\documentclass[12pt]{elsarticle}
\usepackage{graphicx} 
\usepackage{amsfonts} 
\usepackage{amssymb}
\usepackage{enumerate}
\usepackage{url}
\usepackage{rotating}
\usepackage{graphicx}
\usepackage{multicol}
\usepackage{dsfont}
\usepackage{caption}
\usepackage{amsmath, bm}
\usepackage{subcaption}
\usepackage{natbib}
\usepackage[]{algorithm2e}
\usepackage{multirow}
\usepackage{hhline}
\usepackage[export]{adjustbox}
\usepackage{algorithm2e}
\RestyleAlgo{ruled}
\usepackage{float}

\usepackage[toc,page]{appendix}

\usepackage[utf8]{inputenc}
\usepackage{tikz}
\usepackage[bbgreekl]{mathbbol}
\usepackage{ mathrsfs }
\graphicspath{ {./images/} }
\newcommand{\bluecom}[1]{{\color{black} {#1}}}

\begin{document}

\begin{frontmatter}

\title{An FFT based chemo-mechanical framework with fracture: application to mesoscopic electrode degradation}

\author[imdea,upm]{Gabriel Zarzoso}
\ead{gabriel.zarzoso@imdea.es}

\author[uloyola]{Eduardo Roque}
\ead{eroque@uloyola.es}

\author[uloyola]{Francisco Montero-Chacón}
\ead{fpmontero@uloyola.es}

\author[upm,imdea]{Javier Segurado}\corref{cor1}
\ead{javier.segurado@upm.es}

\address[imdea]{IMDEA Materials Institute, 28906, Getafe, Madrid, Spain}

\address[upm]{Departamento de Ciencia de Materiales, Universidad Politécnica de Madrid, C/ Profesor Aranguren 3, 28040 - Madrid, Spain}

\address[uloyola]{Materials and Sustainability, Department of Engineering, Universidad Loyola Andalucía, Avenida de las Universidades s/n, 41704, Dos Hermanas, Seville, Spain.}

\date{\today}

\begin{abstract}
An FFT based method is proposed to simulate chemo-mechanical problems at the microscale including fracture, specially suited to predict crack formation during the intercalation process in batteries. The method involves three fields fully coupled, concentration, deformation gradient and damage. The mechanical problem is set in a finite strain framework and solved using Fourier Galerkin for non-linear problems in finite strains. The damage is modeled with Phase Field Fracture using a stress driving force. This problem is solved in Fourier space using conjugate gradient with an ad-hoc preconditioner. The chemical problem is modeled with the second Fick's law and physically based chemical potentials, is integrated using backward Euler and is solved by Newton-Raphson combined with a conjugate gradient solver. Buffer layers are introduced to break the periodicity and emulate Neumann boundary conditions for incoming mass flux. The framework is validated against Finite Elements the results of both methods are very close in all the cases. Finally, the \bluecom{framework is used to simulate} the fracture of active particles of graphite during ion intercalation. The method is able to solve large problems at a reduced computational cost and reproduces the shape of the cracks observed in real particles.
\end{abstract}


\end{frontmatter}


\section{Introduction}\label{1 Introduction}

Micromechanical models have played a fundamental role in the understanding and design of structural materials in the last decades, helping to understand and predict the effect of the microstructure on the mechanical response of these materials. The improvement of computational power made possible the numerical resolution of these problems by its simulation on representative volume elements (RVEs) of the microstructure. Finite Elements (FE) has been the most common solver used for this type of simulations, but since the seminal work of Moulineq and Suquet \cite{moulinec1994} 30 years ago, Fast Fourier Transform (FFT) based solvers have become a new standard in the field. FFT approaches have been used for solving micromechanical problems in all type of heterogeneous materials including composites, polycrystals, foams and lattice materials. The authors refer to some review articles on the field 
\cite{lucarini2022,schneider2021}. 

Nevertheless, the technological needs have drastically changed in this last 30 years due to the policies to reduce emissions to fight against climate change, which include the reduction of fossil fuels consumption, the rising of renewable energies \cite{energy2020,energy2023} or the use of hydrogen as a fuel or for energy storage. \bluecom{As a result,} new challenges have arise related to the design of materials with optimal mechanical performance for applications as batteries \cite{evreport2023} or hydrogen storage devices. In these cases, the microscopic modeling of the materials involved goes beyond mechanics and requires the simulation of coupled problems, in which mass transport and mechanical equilibrium are present and sometimes combined with damage or temperature. In the case of batteries there is a strong interest in the study of the intercalation process in Li-ion batteries considering its fracture. Different studies have been published recently aimed at simulating the cracking process of active particles in Li ion batteries \cite{lifracture_Pistorio2023}.  In these works, multi-physics coupled problems are solved including the ion diffusion, the deformation of the active particles and the eventual generation of damage involved in the process. The Phase Field Fracture (PFF) model \cite{miehe20101,miehe20102} is the common approach to consider damage in all these studies. Some recent examples in this line include the work by Zuo et.al.  \cite{zuo2015} who modeled the effect of an existing crack in a thin electrode. In \cite{miehe2015battery}, the PFF approach developed by their group for chemo-mechanical problems \citep{miehe20141,miehe20142} is used to simulate the fracture in electrode particles in Li-ion batteries. The method is tested in representative boundary value problems modelling 2D and 3D particle cracking. In \cite{ai2022} PFF is used to simulate the fatigue cracking in Li-ion battery electrode particles. The accumulation of damage through the charge and discharge cycles in a 3D CT scanned particle is studied, and the critical values of C-rate, particle size and initial crack length are determined. Very recently, in \cite{roque2023}, a chemo-mechanical PFF framework based on \cite{miehe2015battery} is used for modelling active particles cracking in dual graphite cathode batteries considering the stochastic nature of graphite fracture.  All the coupled micromechanical problems for batteries reviewed have been solved using finite element frameworks \citep{zuo2015,miehe2015battery,ai2022,roque2023} and the computational cost of the simulations reported was extremely high. The extension of FFT based micromechanical simulation frameworks for the simulation of this type of coupled problems would strongly increase the simulation efficiency, allowing the study of more complex microstructures. 

Some recent studies can be found in the literature which propose FFT based solvers for multiphysical problems with several coupled fields. In \cite{rambausek2018,gokuzum2019} a multiscale framework for magneto-mechanically coupled materials at finite strains is presented. In \cite{schmidt2023} a two-scale method is used to model the thermo-mechanical coupled constitutive behaviour for polycristalline materials, while in \cite{berthensel2019} the thermo-viscoplastic response of composites is studied. 
Regarding problems involving mass transport (chemo-mechanical problems), at least two recent examples can be found. In \cite{chakrabortry2023}, a thermodynamically consistent full field model is integrated within an elasto-viscoplastic FFT framework for the vacancy diffusion effect in plasticity. \cite{pundir2023} propose a framework based in Galerkin FFT to solve corrosion-driven fracture in cementitious materials. However, the chemical processes modeled for vacancy diffusion and corrosion are very different to the intercalation of Li in active particles, and the models developed do not include for example finite strains, fundamental for an accurate description of crack formation in the presence of large ions \cite{roque2023}. No FFT based framework can be found for the simulation of the intercalation process in Li-ion batteries. 

In this paper, an FFT based simulation framework for chemo-mechanical problems with fracture in finite strains is proposed and adapted for solving the cracking of 3D active particles of dual-graphite batteries cathodes. Three different fully coupled problems are involved and solved with FFT methods following an implicit staggered approach \citep{miehe2015battery}. The chemical diffusive problem is governed by the second Fick's law and is integrated with a novel implicit FFT solver based on Krylov subspace methods. The mechanical problem in finite strains is solved with the Fourier Galerkin approach \cite{vondrejc2014, zeman2017, degeus2017} and mixed loading \cite{lucarini20191}. Finally, damage is modeled with the PFF approach and solved with a conjugate gradient method with an ad-hoc preconditioner. The FFT solver has been implemented in code FFTMAD \cite{lucarini20192,lucarini2022}.

The paper is organized as follows. In section 2 the formulation of the different problems and the coupling between them will be explained. In section 3 the FFT solvers for the three individual problems will be presented together resolution of the full coupled problem. In section 4 the FFT implementation is validated against a FE implementation for different study cases and in section 5 the FFT framework will be used in two different numerical examples, a circular 2D particle with stochastic microstructure and a 3D particle with a realistic shape. Finally in section 6 the conclusions of the work will be reviewed.

\section{Chemo mechanical formulation with Phase Field Fracture}\label{2 PFF in chemo-mechanical framework}

\bluecom{The behavior of a Li-ion battery relies on the insertion or extraction of Li ions into particles of active material in the anode, generally made of graphite, during the charge or discharge of the battery. This insertion of ions is named intercalation, and corresponds to a mass transport phenomena where the ions move from the electrolite in contact to the particle surface to the interior of the particles. This intercalation process generates volume changes and stresses in the active material particles, which can eventually lead to their cracking.}

To simulate this process three coupled problems have to be solved involving three different fields, the deformation gradient $\mathbf{F}$ (or the displacement $\mathbf{u}$), the ion concentration $c$ and the phase field representing the fracture, $d$. The model proposed by Miehe et al. \cite{miehe2015battery} including the Phase Field Fracture approach will be the basis of this work, and will be briefly reviewed in this section.

A finite strain formulation is used motivated by the large swelling and shrinking rates experienced by the particles in the case of large ions. The total deformation gradient of a material point is assumed to be multiplicatively decomposed into an elastic and a chemical part
\begin{equation}\label{eq:F=Fe+Fc}
\mathbf{F=F}_e  \mathbf{F}_c.
\end{equation}
The chemical contribution to the deformation, $\mathbf{F}_c$, is due to the volume changes induced by the intercalation of ions in the active particle. This deformation is modeled as a volumetric isotropic expansion depending of the ion concentration $c$ as,
\begin{equation}\label{eq:Fc}
\mathbf{F_c}= (1+\Omega c)^{1/3}\mathbf{I}. 
\end{equation}
where $\Omega$ is a swelling parameter.


\subsection{Mechanical problem and fracture}

A Saint-Venant-Kirchhoff hyperelastic material is assumed for the active particles. The elastic strain measure used is the Green-Lagrange strain 
\begin{equation}\label{eq:Ee}
\textbf{E}_e=\frac{1}{2}(\textbf{F}_e^T \textbf{F}_e-\textbf{I}),
\end{equation}
and the elastic contribution to the free energy of the pristine material corresponds to                   
\begin{equation}\label{eq:psie0}
\psi _{e0} = \frac{\lambda}{2} \mathrm{tr}^2(\textbf{E}_e) + G \mathrm{tr}(\textbf{E}_e^2)  
\end{equation}
where $\lambda$ and $G$ are the Lamé constants.

Fracture is modeled using the Phase Field Fracture (PFF) approach. The  model was introduced in \cite{miehe20101,miehe20102} based on previous works \cite{marigo1998,bourdin2000,bourdin2008} and later extended to finite strains and multiphysics problems in \cite{miehe20141,miehe20142}. The PFF model is based in the replacing the actual sharp crack by a smooth continuous phase field variable $d$
which varies from $d=0$ for the pristine material to $d=1$ for the fully broken state. The phase field is defined through the functional
\begin{equation}
\Gamma(d)=\int_{\Omega} \frac{1}{2{l_c}}d^2 +\frac{l_c}{2}|\nabla d|^2 \mathrm{d}V,
\label{eq:crack}
\end{equation} 
where  $l_c$ is a numerical parameter. A sharp crack is recovered for $l_c\rightarrow0$. 

In the original phase field formulation  \cite{miehe20101} crack evolution is driven by minimization of the total energy functional, considering both the elastic energy and the fracture energy using the crack functional (Eq. \ref{eq:crack}). In the case of materials where fracture is driven by other mechanical field, $d$ is obtained by an alternative evolution equation,
\begin{equation}\label{eq:PFF}
    \frac{g_c}{l_c}\left[d-l^2 \nabla ^2 d \right]= 2(1-d)\mathcal{H}    
\end{equation}
where $g_c$ is the fracture toughness, and $\mathcal{H}$ is the the crack driving force including the history dependence to enforce the irreversibility of the damage process. 

In the case of cracking in graphite particles, fracture is well described using a stress driving force that can be set as the maximum value of an associated crack driving state function $D_f$ as,
\begin{equation}\label{eq:Hist}
\mathcal{H}(\textbf{X} ,t)=\text{max}_{s\in [0,t]} \{ D_f (\textbf{X} ,s)\} \geq 0.
\end{equation}
A principal stress criterion with a threshold is employed here as driving state function $D_f$. Considering the spectral decomposition of the stress in three dimension, $\boldsymbol{\sigma}=\sum_{i=1}^{3}\sigma^i \textbf{n}_i\otimes\textbf{n}_i$ where $\{ \boldsymbol{\sigma} ^i \}_{i=1,3}$ are the principal stresses and $\{ \textbf{n} ^i \}_{i=1,3}$ the eigenvectors, the crack driving state function is defined as 
\bluecom{
\begin{equation}\label{eq:SigDrivingForce}
D_f=\left< \sum_{i=1}^3 \left(\frac{\left<\sigma _i\right>}{\sigma _{max}} \right)^2 -1 \right>
\end{equation}
where $<\cdot>$ indicates positive Macaulay brackets to prevent damage for low or negative stresses and $\sigma _{max}$ is a threshold stress.}

The effect of the phase field in the mechanical response is introduced by degrading the elastic energy of the pristine material $\psi _{e0} $ (Eq. \eqref{eq:psie0}) by a damage prefactor $g(d)=(1-d)^2$. The resulting elastic energy including damage corresponds to
\begin{equation}\label{eq:psiedamaged}
\psi _e = g(d) \psi _{e0} =(1-d)^2\left[ \frac{\lambda}{2} \mathrm{tr}^2(\textbf{E}_e) + G \mathrm{tr}(\textbf{E}_e^2) \right ].
\end{equation}

The balance equation for the mechanical problem is the conservation of linear momentum in finite strains,
\begin{equation}\label{eq:divP}
\nabla \cdot \textbf{P} =0
\end{equation}
where $\textbf{P}$ is the first Piola Kirchoff stress that can be obtained deriving the free energy potential of the system with respect the deformation gradient $\mathbf{F}$ as
\begin{equation}\label{eq:Pderivada}
\textbf{P}=   \frac{\partial \psi _e}{\partial \textbf{F}}=(1-d)^2 \frac{\partial \psi _{e0}}{\partial \textbf{F}}=(1-d)^2\textbf{P}_0(\textbf{E}_e).
\end{equation} 
The function $\textbf{P}_0$ corresponds to the stress in a pristine material, and is given by
\begin{equation}\label{eq:pristine}
\textbf{P}_0 = 
\textbf{F} (\lambda \mathrm{tr}(\textbf{E}_e)\textbf{I}+2G\textbf{E}_e).
\end{equation} 

\subsection{Chemical problem}
The chemical problem consist in finding the evolution of the local concentration of ions  as function of time, $c(\mathbf{x},t)$. The continuity equation for this concentration is modeled with a second Fick's law as
\begin{equation}\label{eq:DifOriginal}
\frac{\partial c}{\partial t}+ \nabla \cdot \textbf{J}_c = \bluecom{\dot{c}_{source}}
\end{equation}
where $\textbf{J}_c$ is the mass flow and $\dot{c}_{source}$ is a volumetric source term.

The mass flow  $\textbf{J}_c$ is derived from the bulk energy following non-equilibrium thermodynamics statements and includes the effect of damage.  The starting point for obtaining $\textbf{J}_c$ is the chemical potential $\mu$, which corresponds to the derivative of the free energy potential with respect $c$
\begin{equation}\label{eq:mupartial}
\mu = \frac{\partial \psi}{\partial c}.
\end{equation}

The free energy density is split in two terms,
\begin{equation}\label{eq:psitotal}
\psi(\textbf{F},c,d) = \psi _{e}(\textbf{F},c) + \psi _c(c),
\end{equation}
the mechanical energy $\psi _{e}$ (Eq.\eqref{eq:psiedamaged}) and the chemical free energy $\psi _{c}$, given by
\begin{equation}\label{eq:psic}
\psi _c(c) =RT c \log(c)+RT(1-c)\log(1-c)
\end{equation}
being $T$ the temperature and $R$ the universal gas constant.

The differentiation of the free energy function give rise to two contributions to the energy potential, the elastic $\mu_e$ and the chemical $\mu_c$ ones. The chemical contribution $\mu_c$ is obtained deriving expression \eqref{eq:psic} with respect to $c$, leading to
\begin{equation}\label{eq:muc}
\mu _c= \frac{\partial \psi _c}{\partial c} =RT \log \left( \frac{c}{1-c}\right).
\end{equation}
The elastic contribution to the chemical potential results in 

\begin{equation}\label{eq:mue}
\mu _e= \frac{\partial \psi _e}{\partial c}= \frac{\Omega}{3}(\lambda+2G)\left(-2(1+\Omega c)^{-7/3}\mathrm{tr}(\mathbf{C})  + (1+\Omega c)^{-5/3}
2\mathrm{tr}(\mathbf{C})\right)
\end{equation}

with $\mathbf{C}$ the right Cauchy–Green deformation tensor, $\mathbf{C}=\mathbf{F}^{T}\mathbf{F}$.
\noindent A  dissipation potential $\phi$ with quadratic dependence on \bluecom{the total energy potential $\mu=\mu_e+\mu_c$} is assumed, 
\begin{equation}\label{eq:phidissipationpotential}
\phi= \frac{M(d)}{2} c(1-c)\textbf{C}^{-1} : \nabla \mu \otimes \nabla \mu  
\end{equation} 
 being $M$ the mobility. In \cite{miehe2015battery} a degradation of the mobility is proposed which reflects the effect of microporosity in the transport of ions, and is given as
\begin{equation}\label{eq:mobility}
M(d)=-\frac{D}{RT} (0.9(1-d)^2 +0.1d^2)
\end{equation} 
being $D$ the diffusivity of the ions, $R$ the universal gas constant and $T$ the absolute temperature. Using this dissipation potential, the mass flow $\textbf{J} _c$ is given by
\begin{equation}\label{eq:Jmassflowpartial}
\mathbf{J}_c=\frac{\partial \phi}{\partial \nabla \mu}	 
\end{equation}
obtaining
\begin{equation}\label{eq:Jmassflowexpression}
\mathbf{J}_c={M(d)}c(1-c)\textbf{C}^{-1} \nabla \mu.
\end{equation}
\bluecom{where the mobility function $M(d)$ is given in Eq. \eqref{eq:mobility}, $\textbf{C}^{-1}$ is the inverse of Cauchy-Green deformation tensor and the total potential $\mu$ is given by the elastic and chemical contributions as $\mu=\mu_e+\mu_c$}.

\section{FFT based simulation framework}\label{3 FFT based framework}
Mechanical equilibrium and damage equations will be solved following an implicit scheme, forcing their balance equations to be fulfilled with the value of the fields at the current time step. To alleviate the notation, the time in which the fields are defined will be indicated as subindex, e.g. $\mathbf{F(x},t)=\mathbf{F}_t$. The conservation of mass for the chemical problem (Eq. \ref{eq:DifOriginal}) is a \bluecom{parabolic} equation and therefore has to be integrated in time.

\noindent
\bluecom{\subsection{Microscopic problems and homogenization}}

The resulting system of coupled partial differential equations to be solved includes Eq. \eqref{eq:divP}, that stands for mechanical equilibrium, Eq. \eqref{eq:DifOriginal} for the balance of ion concentration and Eq. \eqref{eq:PFF} for the Phase Field Fracture problem. In the context of mesoscopic/microscopic problems, these equations are solved at the lower length scale to provide the micro-field distribution of the deformation gradient $\textbf{F}: [\Omega,t]\rightarrow \mathbb{R}^{3\times3}$, $c: [\Omega,t]\rightarrow \mathbb{R}$ and $d:[\Omega,t]\rightarrow \mathbb{R}$. Regarding boundary conditions and constraints, all the three problems are solved assuming periodic boundary conditions, which fulfill the Hill-Mandel relation for the mechanical and chemical problem.

In the mechanical problem, the macroscopic value of the deformation gradient $\overline{\textbf{F}}(t)$ is prescribed as function of time. The microscopic deformation gradient can then be decomposed as
\begin{equation} \label{eq:Fsplit}
\textbf{F(x},t)=\overline{\textbf{F}}(t)+\tilde{\textbf{F}}(\mathbf{x},t)
\end{equation}
where $\tilde{\textbf{F}}$ is the deformation gradient fluctuation, which is periodic, and the average of $\textbf{F(x},t)$ fulfills
\begin{equation}
<\textbf{F(x},t)>_\Omega = \overline{\textbf{F}}(t).
\end{equation}
In this form, in the mechanical problem the variable to be solved at each time step is the deformation gradient fluctuation $\tilde{\textbf{F}}$ which is fully periodic.

In the chemical problem, the concentration of ions $c$ is the primary variable and fulfills periodic boundary conditions at each time step. In this case, no special macroscopic conditions are needed, and the total ion mass and average concentration are a natural consequence of the mass introduced in the source term of Eq. \eqref{eq:DifOriginal}, leading to
\begin{equation}
<c(\mathbf{x},t+\Delta t)>_\Omega = <c(\mathbf{x},t)>_\Omega + <\dot{c}_{B}(\mathbf{x},t+\Delta t)>\Delta t
\end{equation}
where implicit integration is assumed.

Finally, in the phase field problem the variable solved is directly $d$ and therefore is fully periodic. Its total value is not prescribed and evolves naturally following Eq. \eqref{eq:PFF}.

\noindent
\bluecom{\subsection{FFT resolution algorithms}}

In order to solve the governing equations presented in section \ref{3 FFT based framework} using Fourier based algorithms, the first step is to discretize the simulation domain in voxels and the Fourier space in the corresponding set of discrete frequencies. The simulation domain is a periodic RVE shaped as a cube of length $L_x$,$L_y$,$L_z$, discretized in a regular array of $N_x$,$N_y$,$N_z$ voxels. 

The functions discretized in real space as the value of the field on the center of each voxel are transformed to Fourier space using the Discrete Fourier Transform (DFT), providing a discrete map with values corresponding to each of the $N_x\times N_y \times N_z$ frequencies. In order to avoid Gibbs oscillations and aliasing effects when solving heterogeneous materials with high contrasts, modified frequencies are used based on expressing continuum derivatives in terms of finite differences  \citep{willot2015} .

\subsubsection*{Mechanical Equilibrium}
The mechanical problem (Eq. \ref{eq:divP}) is resolved using the Fourier Galerkin approach in FFT \cite{vondrejc2014} on a finite deformation framework \cite{zeman2017}. This approach starts by formulating the linear momentum conservation Eq.(\ref{eq:divP}) in a weak form derived from the principle of virtual work. \bluecom{After discretization of the fields in trigonometrical polynomials and integration, the resulting system of equations is equivalent to the one derived transforming the Lippmann Schwinger equation in the original basic scheme into a linear system, as shown in \cite{zeman2010,vondrejc2014}. Therefore, weak and strong formulations of the homogenization problem are equivalent in FFT.} In a finite deformation framework, the equilibrium equation reads
\begin{equation}\label{eq:virtualworkP}
\int _{\Omega} \zeta(\mathbf{x}) : \textbf{P} (\textbf{F}(\mathbf{x}),c(\mathbf{x}),d(\mathbf{x}))\mathrm{d}\Omega = \textbf{0}
\end{equation}
where $\zeta(x) $ is an arbitrary test function. In this c problem the value of $\textbf{P}$ depends on other two fields, the damage $d(\mathbf{x})$ as stated in Eq. \eqref{eq:Pderivada} and $c(\mathbf{x})$ due to the presence of stress-free chemical strain, 
\begin{equation}\label{eq:PFcd}
\textbf{P}(\mathbf{F},c,d) = (1-d)^2 \textbf{P}_0\left( \mathbf{F}_e \right) =  (1-d)^2 \textbf{P}_0\left( (1+\Omega c)^{-1/3} \textbf{F}\right) 
\end{equation}
where $\textbf{P}_0$ is the stress in a pristine material, given by Eq. \eqref{eq:pristine} in the case an elastic Saint Venant-Kirchoff material. Both fields $c$ and $d$ influence the local mechanical behavior inducing an heterogeneous response even for a single phase microstructure. If deformation gradient is split in its average and fluctuating parts (Eq. \ref{eq:Fsplit}), \bluecom{the problem consist in finding the deformation gradient fluctuation, $\tilde{\mathbf{F}}$ which fulfills Eq. \eqref{eq:virtualworkP} for any compatible virtual field $\zeta$ and a prescribed macroscopic deformation gradient $\overline{\mathbf{F}}$. In general, the full macroscopic deformation gradient can be prescribed as function of time $\overline{\mathbf{F}}(t)$, or can be combined with macroscopic stress conditions as described in \cite{lucarini20191}}

The compatibility of the field $\zeta$ is imposed using a projection operator which maps any arbitrary periodic field $\xi$, on its compatible part by a convolution operation 
\begin{equation}\label{eq:mapping G}
\zeta(\mathbf{x}) =  \mathbb{G} \ast \xi (\mathbf{x})
\end{equation}
where the projection operator $\mathbb{G}$ is equivalent to the double derivative of Green' s function as explained in \cite{degeus2017}.

Introducing \eqref{eq:mapping G} in the weak form of equilibrium \eqref{eq:virtualworkP} results in
\begin{equation}\label{eq:integral G}
\int _{\Omega} \mathbf{\xi}(\mathbf{x}): \left[\mathbb{G}\ast \textbf{P}(\mathbf{x})\right]\mathrm{d} \Omega = 0.
\end{equation}

The method approximates the fields involved in the integral using trigonometrical polynomial interpolation by a collocation method which identifies the coefficients of the polynomials with a discrete Fourier transform. After discretization, a trapezoidal rule is used for integration. The resulting expression is a non-linear system of algebraic equations that provides the value of the discrete finite strain fluctuations $\tilde{ \textbf{F}}$ at the grid points,

\begin{equation}\label{eq:mechanicalsolve}
    \mathcal{F}^{-1}\left[\widehat{\mathbb{G}}:\mathcal{F}[\mathbb{K} : \tilde{\textbf{F}}] \right]=-\mathcal{F}^{-1}\left[\widehat{\mathbb{G}}:\mathcal{F}[\textbf{P}(\textbf{F})] \right]
\end{equation}
where $\mathbb{K}$ stands for the tangent stiffness of the material response at each point,
$$
\mathbb{K}(\mathbf{x})=\frac{\partial \mathbf{P}}{\partial \mathbf{F}}.
$$
The non-linear equation \eqref{eq:mechanicalsolve} is solved by linearization using Newton-Raphson and, for each linear iteration, a Krylov solver is used for the linear system, here the conjugate gradient method. The solution of a Newton iteration $i$ is the correction $\delta\mathbf{F}$, which has zero average. The deformation gradient at the end of the procedure is the sum of the macroscopic strain applied, and all the corrections. \bluecom{To stop the Newton iterations several criteria can be used and here, as usually done in finite elements, an equilibirium criterion is chosen. Since a tensor field $\boldsymbol{\sigma}$ has zero divergence if $\mathbb{G}\ast \boldsymbol{\sigma} =0$ \citep{moulinec1998}, this value properly normalized is chosen as residual. The expression is,
\begin{equation}\label{mechresiduum}
err_{res}= \parallel \mathcal{F}^{-1}\left[\widehat{\mathbb{G}}:\mathcal{F}[\textbf{P}( \textbf{F})] \right] \parallel /\parallel \mathcal{F}^{-1}\left[\widehat{\mathbb{G}}:\mathcal{F}[\textbf{P}(\overline{\textbf{F}})] \right] \parallel
\end{equation}
where $\textbf{P}(\textbf{F})$ is the actual stress and $\textbf{P}(\overline{\textbf{F}})$ is the comparison stress for normalization.
} In the mechanical problem, the rotated differentiation scheme in \cite{willot2008} is used.  The algorithm solving the mechanical equilibrium is shown in  algorithm \ref{alg:Mech}.

\vspace{5mm}
\begin{algorithm}[H]
\caption{\bluecom{Mechanical problem}}\label{alg:Mech}
\KwData{$\Delta\overline{\textbf{F}}_{t+\Delta t}, c(\mathbf{x}),d(\mathbf{x})$}
\KwResult{$\textbf{F}(\textbf{x})$}
$\textbf{F}^{i=0}=\textbf{F}_t+\Delta \bar{\textbf{F}}_{t+\Delta t}$\; 
\While{$\bluecom{err_{res} } > tol_{nw}$}{
  $\textbf{P}=\textbf{P}(\textbf{F}^i,c,d)$ (Eq. \eqref{eq:PFcd})\;
  $\mathbb{K}=\frac{\partial \textbf{P}}{\partial \textbf{F}}|_{\textbf{F}=\textbf{F}^i}$\;
  Find $\delta \textbf{F}$ by Conjugate Gradient with a tolerance of $tol_{lin}$ for:
        $\mathcal{F}^{-1}\left[\widehat{\mathbb{G}}:\mathcal{F}[\mathbb{K} : \delta \textbf{F}] \right]=-\mathcal{F}^{-1}\left[\widehat{\mathbb{G}}:\mathcal{F}[\textbf{P}] \right]$\
        
  $\textbf{F}^{i+1}=\textbf{F}^i+\delta\textbf{F}$\
}
$\textbf{F}_{t+\Delta t}=\textbf{F}^{i+1}$
\end{algorithm}

\vspace{5mm}
\subsubsection*{Fracture problem}
The objective is obtaining the fracture phase field $d(\mathbf{x})$ \bluecom{for a given driving force $\mathcal{H}$}, which depends on the current value of the deformation gradient, by solving the phase field equation (\ref{eq:PFF}). This equation is a non-homogeneous Helmholtz equation with the source term given by $\mathcal{H}(\mathbf{x})$, and linear in $d$. Rearranging terms leads to,
\begin{equation}\label{eq:PFFrearranged}
\left[\frac{g_c}{l_c} + 2\mathcal{H}\right]d-g_c l_c \Delta d = 2 \mathcal{H}.
\end{equation}

In order to solve Eq. \ref{eq:PFFrearranged}, it is transformed to Fourier space and discretized in voxels. The resulting equation, transformed back to real space reads
\begin{equation}\label{eq:PFF_Fourier}
\left[\frac{g_c}{l_c} + 2\mathcal{H}\right]d + \mathcal{F}^{-1} \left\{ g_c l_c \  \boldsymbol{\xi}\cdot \boldsymbol{\xi}  \hat{d} \right\}  = 2 \mathcal{H}.
\end{equation}
This equation corresponds to a linear map $\mathcal{A}()$ acting on the discrete field $d: \mathbb{R}^{N_x}\times \mathbb{R}^{N_y}\times \mathbb{R}^{N_z}\rightarrow \mathbb{R}^{N_x}\times \mathbb{R}^{N_y}\times \mathbb{R}^{N_z}$ equaled to an independent term, $ \mathcal{A}(d)=b$, with
\begin{equation}\label{eq:PFFA}
\mathcal{A}(d)= \left[\frac{g_c}{l_c} + 2\mathcal{H}\right]d + \mathcal{F}^{-1} \left\{ g_c l_c \  \boldsymbol{\xi}\cdot \boldsymbol{\xi}  \hat{d} \right\}
\end{equation}
and
\begin{equation}\label{eq:PFFb}
b=2\mathcal{H}.
\end{equation}
This linear system of equations is symmetric positive definite and can be solved using conjugate gradient, which does not require to compute and store the linear operator as a matrix. For an efficient resolution of the system is fundamental having a good preconditioner. Following a similar derivation than \cite{lucarini20193} in the case of mechanical equilibirum equation or \cite{sancho2023} for elasto-dynamics (which once integrated results the same Helmholtz operator) a linear operator preconditioner $\mathcal{P}$ reads
\begin{equation}\label{eq:PFFprecond}
\mathcal{P} \approx \mathcal{A}^{-1}(b(\mathbf{x}))= \mathcal{F}^{-1}\left\{ \frac{1}{g_c/l_c + 2\overline{\mathcal{H}}+\boldsymbol{\xi}\cdot \boldsymbol{\xi} g_cl_c}\widehat{b}(\boldsymbol{\xi})\right\}
\end{equation}
where $\overline{\mathcal{H}}$ is the average value of the history.
\noindent The scheme for solving the phase field problem is reviewed in algorithm \ref{alg:PFF}.

\vspace{5mm}
\begin{algorithm}[H]
\caption{\bluecom{Phase Field Fracture problem}}\label{alg:PFF}
\KwData{$\mathcal{H(\textbf{x})},tol_{cg}$}
\KwResult{$d(\mathbf{x})$}
Solve $d(\mathbf{x})$ by Conjugate Gradient with a tolerance of $tol_{cg}$ for:\
    $\left[g_c/l_c + 2\mathcal{H}\right]+g_c l_c \mathcal{F}^{-1} \left[\boldsymbol{\xi}\cdot \boldsymbol{\xi} \mathcal{F}[d]\right]=2\mathcal{H}$ \

with preconditioner:
$\mathcal{P} (\cdot)= \mathcal{F}^{-1}\left\{ \frac{1}{g_c/l_c + 2\overline{\mathcal{H}}+\boldsymbol{\xi}\cdot \boldsymbol{\xi} g_cl_c}\mathcal{F}(\cdot)\right\}$
\end{algorithm}

\vspace{5mm}
\subsection*{Chemical problem}\label{sec:ChemProblemResolution}

For simulating ion transport in active particles immersed in an electrolyte, in addition to the standard chemo-mechanical equations described, it is necessary to introduce a variable boundary condition that emulates the solution of a Butler-Volman equation \cite{miehe2015battery} in the boundary between a particle and the surrounding electrolyte. In a periodic problem, active particles are fully embedded in the simulation domain and boundary terms cannot be imposed directly. As an approximate alternative, the inner ion flow from electrolyte to active particle, $\mathbf{q}_{in}$ is modeled introducing an additional term $\dot{c}_{B}$ in the equation \eqref{eq:DifOriginal},

\begin{equation}\label{eq:DifCCterm}
\frac{\partial c}{\partial t}+ \nabla \cdot \mathbf{J}_c = \dot{c}_{source} + \dot{c}_{B}.
\end{equation}
In Eq.\ref{eq:DifCCterm}, the volume source $\dot{c}_{B}$ fulfills
\begin{equation}\label{eq:neumann1}
\int_\Gamma \mathbf{q}_{in}\cdot \mathbf{n} \ \mathrm{d} \Gamma = \int_\Omega \dot{c}_{B} \delta_\Gamma(\mathbf{x}) \mathrm{d} \Omega
\end{equation}
where $\delta_\Gamma(\mathbf{x})$ is a surface delta function 
\begin{equation}\label{eq:delta}
\delta_\Gamma(\mathbf{x})= \left\{ \begin{array}{c} 
0 \ \text{if}\ \mathbf{x}\in \Gamma \\
\infty \ \text{if}\ \mathbf{x}\notin \Gamma 
\end{array}\right.
\end{equation}
From  a practical view point,  $\delta_\Gamma(\mathbf{x})$ is taken as a narrow region with size $\delta L$ near the boundary with value 1. In this case, for each point of the boundary the value of $\dot{c}_B$ in the attached internal region defined by $\delta(\mathbf{x})$ corresponds to
\begin{equation}\label{eq:neumann2}
\dot{c}_B=\mathbf{q}_{in}\cdot \mathbf{n} /\delta L.
\end{equation}

\bluecom{In the chemical problem is necessary to solve the equation \eqref{eq:DifCCterm} with this flux $\dot{c}_{B}$ imposed. To solve Eq. \eqref{eq:DifCCterm}, first, the equation is discretized in time, in this case following a Backward-Euler approach.} Considering a regular time discretization with time increments $\Delta t$, the resulting equation becomes
\begin{equation}
\label{eq:chem_int}
\frac{c_{t+\Delta t}-c_{t}}{\Delta t}+ \nabla \cdot \textbf{J}_c(\mathbf{F},c_{t+\Delta t},d) = \dot{c}_{source}(t+\Delta t) + \dot{c}_{B}(t+\Delta t).
\end{equation}

To simplify the resolution, a semi-implicit integration is proposed where the term $c(1-c)$ multiplying $\nabla \mu$ in Eq. \eqref{eq:phidissipationpotential} is evaluated with the concentration of the previous step as $c_t(1-c_t)$. The resulting equation to be solved at $t+\Delta t$ remains implicit due to the dependence in $c$ of the chemical potential $\mu(c)$. \bluecom{In the of absence of an additional source of concentration, $\dot{c}_{source}(t+\Delta t)=0$, the resulting equation is}
\begin{equation}\label{eq:ChemDTsolve}
\frac{c-c_t}{\Delta t} + \nabla \cdot \left[M(d_t)c_t(1-c_t) \textbf{C}_t^{-1}\nabla \mu(c,\mathbf{F}_t) \right]=\dot{c}_{B}.
\end{equation}

The resolution of this equation is done using a Newton-Raphson scheme, solving in an iterative manner a linearized version of \eqref{eq:ChemDTsolve}.
Let $f(c)$ be the residual of
\begin{equation}\label{eq:Chemresidualfc}
 f(c)= c + \nabla \cdot \left[M(d_t)c_t(1-c_t) \textbf{C}^{-1}\nabla \mu \right] \Delta t - \dot{c}_{B}  \Delta t - c_t 
\end{equation}
then, the value of $f$ can be linearized around a reference concentration $c^i$ as,
\begin{equation}\label{eq:Chemresidualfcref}\left.
f(c^i+\delta c )\approx f(c^i)+\frac{\partial f}{\partial c}  \right| _{c^i}  \delta c .
\end{equation}
The value of $f(c^i)$ is obtained by substituting $c^i$ on Eq. \eqref{eq:Chemresidualfc} where $\mu(c)$ is obtained from Eqs. \eqref{eq:muc} and \eqref{eq:mue}. The derivative of the residual acting on a perturbation concentration field $\delta c$ corresponds to
\begin{equation}\label{eq:Chemderfc}
\left. \frac{\partial f}{\partial c} \right|_{c^i} \delta c =\delta c -\Delta t\nabla \cdot \left[  M(d_t) c_t(1-c_t)\textbf{C}^{-1} \nabla\left[ \frac{1}{c^i}+\frac{1}{1-c^i}\right]\delta c \right]
\end{equation}

With these expressions, at each Newton iteration by setting $f(c^i+\delta c)=0$, \bluecom{a linear problem is derived to obtain the perturbation $\delta c$,}
\begin{equation}\label{eq:newton}
\left[ \left. \frac{\partial f}{\partial c}  \right|_{c^i} \right]\delta c =-f(c^i)
\end{equation}

The Newton-Raphson scheme is initiated in $c^0=c_t$ and, in each iteration, $\delta c $ will be obtained solving the linear system \eqref{eq:newton}. This increment $\delta c $ is added to the input concentration $c^i$ as
\begin{equation}\label{eq:Chemcref+dc}
c^{i+1} =c^{i}+\delta c.
\end{equation}
This is done iteratively until $\delta c$ is small enough respect to the initial concentration 
\begin{equation}\label{eq:Chemerror}
\parallel \delta c \parallel / \parallel c_t \parallel < tol_{chem}
\end{equation}
and the Newton-Raphson scheme ends. The resolution of the linear problem (Eq. \eqref{eq:newton}) is done by transforming the equation to Fourier space and using the conjugate gradient. \bluecom{The frequencies used are defined assuming a forward-and-backward finite difference scheme \cite{willot2015}}. The Fourier transform of the operator in Eq. \eqref{eq:newton}  corresponds to

\begin{equation}\label{eq:ChemresidualFourier_A}
   \left[ \widehat{  \left. \frac{\partial f}{\partial c}  \right|_{c^i}} \right]\widehat{\delta c} = \widehat{\delta c} -\Delta t   \left\{\boldsymbol{\xi}\cdot \mathcal{F} \left\{  M(d_t) c_t(1-c_t)\textbf{C}^{-1} \mathcal{F}^{-1} \left\{\boldsymbol{\xi} \mathcal{F} \left\{ \frac{1}{c^i}+\frac{1}{1-c^i}\right\}\right\}\widehat{\delta c}\right\}\right\},
\end{equation}
while the independent term in the right hand side transformed to Fourier space is 
\begin{equation}\label{eq:ChemresidualFourier}
    \widehat{f(c^i)}= \widehat{c^i} + \Delta t\left\{\boldsymbol{\xi}\cdot \mathcal{F} \left\{M(d_t)c_t(1-c_t) \textbf{C}^{-1} \mathcal{F}^{-1}\left\{\boldsymbol{\xi} \hat{\mu}(c^i) \right\}\right\} \right\} -  \Delta t\widehat{\dot{c}_{B}} - \hat{c_t}.
\end{equation}
\noindent The process to solve the chemical problem is illustrated in Alg.\ref{alg:Diff}.

\vspace{5mm}
\begin{algorithm}[H]
\caption{\bluecom{Chemical diffusion problem}}\label{alg:Diff}
\KwData{$c_t(x)$, $\textbf{F}$, $\dot{c}_{B}$, $\Delta t$, $tol_{chem}$, $tol_{lin_{chem}}$}
\KwResult{$c_{t+\Delta t}(x)$}
$c^i = c^t$\;
\While{$\parallel \delta c\parallel / \parallel c^t \parallel > tol_{chem}$}{
  Solve $\delta c $ by Conjugate Gradient with a tolerance of $tol_{lin_{chem}}$ for:
    $\left[\frac{\partial f}{\partial c}|_{c^i} \right]\delta c =-f(c^i,\dot{c}_{B})$
    
  $c^{i+1} = c^i+ \delta c $\

  $c^i=c^{i+1}$
}
$c_{t+\Delta t}=c^{i+1}$
\end{algorithm}

\vspace{5mm}
\subsubsection*{Global problem}
For each time step, the resolution of the equations is done in a staggered way, solving the problems sequentially and, for each equation assuming as constant the fields not involved directly in the corresponding balance. The sequential resolution of the equations is done until convergence when the three fields are in equilibrium (implicit staggered). An exception to this is the chemical problem, which is solved only once for each time step, based on the values of deformation gradient and damage of the previous time step. This approximation is motivated by the small influence of these variables in the chemical potential.

First, the variables are set for time $t=0$, considering an undeformed state, $\textbf{F}_{0}(\mathbf{x})=\mathbf{I}$ and no initial damage, $d_0(\mathbf{x})=0$. Regarding the chemical problem, an homogeneous initial concentration of ions $c(\mathbf{x})=c_{0_{min}}=10^{-2}$ is set to avoid problems of convergence due to the logarithms in expression (\ref{eq:muc}). 

After the resolution of a time step $t$, the three fields $\mathbf{F}_t,c_t,d_t$ which fulfield the Eqs. (\ref{eq:divP},\ref{eq:chem_int},\ref{eq:PFF}) are stored. To obtain the solution for the next time step ($t=t+\Delta t$), the applied macroscopic deformation gradient is updated from its value in the previous step $\textbf{F}_{t+\Delta t}=\overline{\textbf{F}}_{t+\Delta t} + \tilde{\textbf{F}}_{t}$. The damage for the new time step $t+\Delta t$ is just taken as its value on the previous time, $d_{t+\Delta t}=d_{t}$

Then, the chemical problem (Eq. \eqref{eq:chem_int}) is solved considering the new macroscopic deformation gradient and the value of damage of previous time step. The resolution of this non-linear problem provides a new distribution of concentration $c_{t+\Delta t}$ that will be subsequently used in the mechanical problem with damage. These two problems are solved in a staggered implicit manner. Let $k$ be last staggered iteration where both $d^k$ and $c^k$ have been obtained. The mechanical problem for a new iteration will be solved to achieve stress equilibrium obtaining a new deformation gradient $\textbf{F}_{t+\Delta t}^{k+1}$, using damage and concentration, $d^k$ and $c^k$, of previous iteration. The damage problem is solved then with the new stress $\mathbf{P}^{k+1}_{t+\Delta t}$ acting as driving force, to provide a new iteration of the phase field $\textbf{d}_{t+\Delta t}^{k+1}$. The staggered iterations of mechanical and damage problems stop when the problem reaches mechanical equilibrium, using a residual  defined coherently with the FFT implementation of the mechanical problem. After convergence, the three fields, $c$, $\textbf{F}$ and $d$ are in equilibrium and the step is completed. 
The detailed algorithm of the process is shown in algorithm \ref{alg:Staggered chemo}.

\vspace{5mm}
\begin{algorithm}[H]
\caption{\bluecom{Global Staggered scheme }}\label{alg:Staggered chemo}
\KwData{$\overline{\textbf{F}}(t),\dot{c}_{B}(t),\Delta t,t_{end},tol_{lin},tol_{nw},tol_{chem},tol_{st}$}
\KwResult{$c(\textbf{x},t),\textbf{F}(\textbf{x},t),d(\mathbf{x},t) $}
$t=0$\

$c_{t}=c_{min}$\

$\textbf{F}^{k=0}_{t}=\textbf{I}$

$d_{t}^{k=0}=0$\

\While{$t\leq t_{end}$}{
  Solve chemical problem (alg:\ref{alg:Diff}) $\longrightarrow$ $c_{t+\Delta t}$\; 
    \While{$err_{res} > tol_{mech}$}{
      
      calculate $err_{res}$ as \bluecom{\eqref{mechresiduum}}
    
      Solve mechanical problem (alg:\ref{alg:Mech}) $\longrightarrow$ $\textbf{F}_{t+\Delta t}^{k+1}$\;
      
      Solve PFF problem (alg:\ref{alg:PFF}) $\longrightarrow$ $d_{t+\Delta t}^{k+1}$\;

      $\textbf{F}_{t+\Delta t}^{k}=\textbf{F}_{t+\Delta t}^{k+1}, d_{t+\Delta t}^{k}=d_{t+\Delta t}^{k+1}$ 
      
  }
  $\textbf{F}_{t+\Delta t}=\textbf{F}_{t+\Delta t}^{k}$\;
  $d_{t+\Delta t}=d_{t+\Delta t}^{k}$\;
  $t=t+\Delta t$\;
}
\end{algorithm}

\section{Validation of the FFT framework with FE}
The results of the FFT framework developed for the chemo-mechanical problem with fracture is compared with an equivalent implementation for the Finite Element method using the open source free software FEniCS. In all the cases the finite element domain is discretized in a regular mesh of triangular linear elements. Further details of the FE implementation are described in \cite{roque2023}. 

\subsection{Neumann Boundary Conditions}
The first test is used to validate the introduction of the source term in the FFT formulation, Eq. \eqref{eq:DifCCterm} to emulate a Neumann boundary condition. The domain used for this test is a 3D thin plate of length $L_x=0.010 \mu$m, $L_y=0.010 \mu$m, $L_z=1.5625\ 10^{-4}\mu$m discretized in $N_x=64$, $N_y=64$ voxels with one voxel thickness $N_z=1$.  In FE the equivalent discretization consists of 8192 triagnular elements.

\begin{figure}[H]
    \centering
    \begin{subfigure}{0.45\textwidth}
		\includegraphics[width=\textwidth]{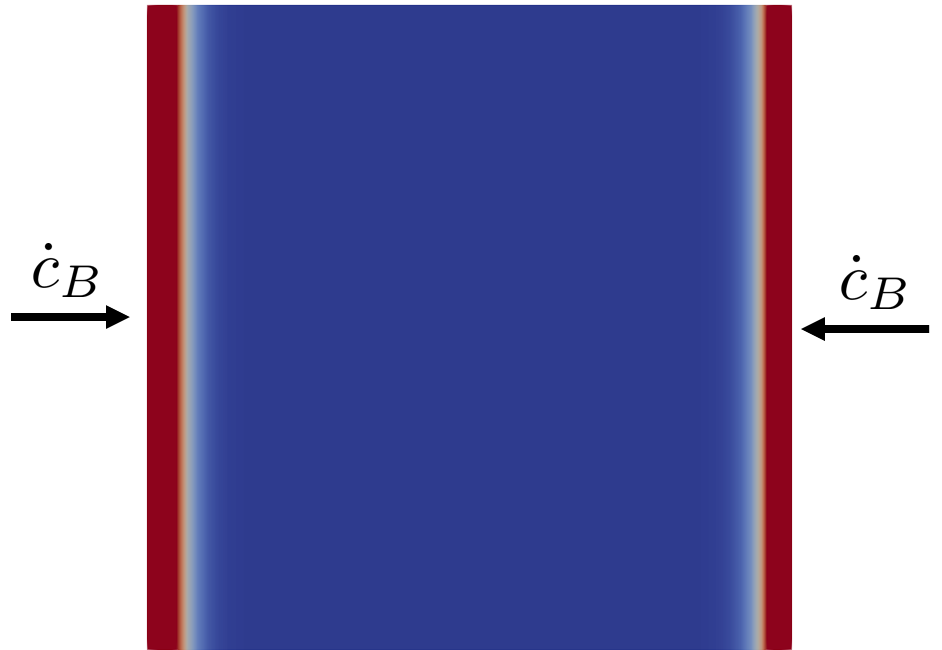}
        \includegraphics[width=\linewidth]{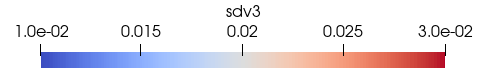}    
    \end{subfigure}
    \begin{subfigure}{0.45\textwidth}
		\includegraphics[width=\textwidth]{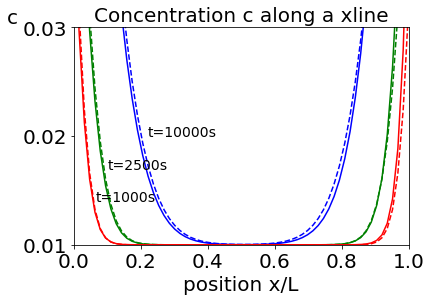}
        \centering
        \includegraphics[width=0.5\linewidth]{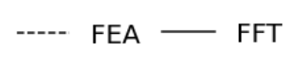}  
	\end{subfigure}
       \caption{Concentration $c$ distribution with Neumann boundary conditions in FE vs FFT. On the left the concentration at time $t= 1000 (s)$ is represented. A flux is set at the lateral boundaries and concentration diffuses to the inner part. On the right the concentration along a horizontal line of the domains at times $t= 1000 (s)$ (red),  $t= 2500 (s)$ (green) and  $t= 10000 (s)$ (blue) is plotted for both FE and FFT frameworks. }
       \label{fig:4.1_dibujo} 
\end{figure}

The initial concentration is $c_0=10^{-2}$ and the diffusivity $D=10^{-9}$mm$^2$/s. The Neumann boundary condition in Finite Elements corresponds to an incoming flow on the horizontal boundaries of value $\mathbf{q}\cdot \mathbf{n}=J_{in}=10^{-6}$ mol/mm$^2$s$^{-1}$ (Fig. \ref{fig:4.1_dibujo}). To emulate this flux in FFT a source term is introduced which fulfils Eqs. \eqref{eq:neumann1} and \eqref{eq:neumann2}. In this case, one voxel is used as the size $\delta L$ of the delta function (Eq. \ref{eq:delta}), being $\dot{c}_{B}=6.5 \ 10^{-7}$ mol/mm$^3$ s$^{-1}$ in the voxels of the region defined by $\delta(\mathbf{x})$.

In  the same figure the concentration obtained in FFT for $t=1100$s is represented as coloured iso-plot and it can be observed how the solution only depends on the horizontal position $x$. The concentration along a horizontal line at different times obtained both using FE and FFT is illustrated in Fig. \ref{fig:4.1_dibujo}. Differences are very small with an average of $0.047 \%$ at $t=1000s$, $0.075 \%$ at $t=2500s$ and $0.14 \%$ at $t=10000s$.

\begin{figure}[H]
	\centering
	\begin{subfigure}{0.32\textwidth}
		\includegraphics[width=\textwidth]{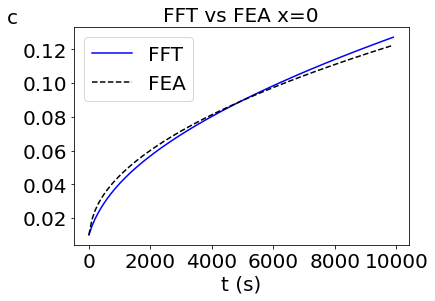}
		\caption{$c$ in $x=0.0$}
		\label{fig:4.1_grafica_x00}
	\end{subfigure}
	\begin{subfigure}{0.32\textwidth}
		\includegraphics[width=\textwidth]{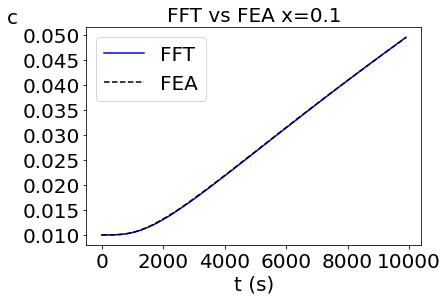}
		\caption{$c$ in $x=0.1l$}
		\label{fig:4.1_grafica_x01}
	\end{subfigure}	
	\begin{subfigure}{0.32\textwidth}
		\includegraphics[width=\textwidth]{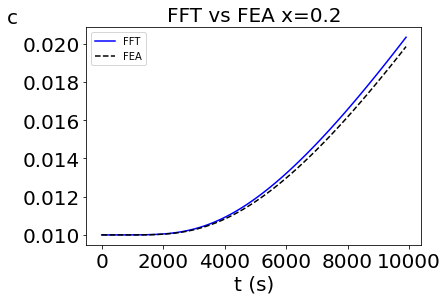}
		\caption{$c$ in $x=0.2l$}
		\label{fig:4.1_grafica_x02}
	\end{subfigure}	
	\caption{Concentration values at different points using Neumann boundary conditions in FEnics and source term $\dot{c}_{B}$ in FFT.}
	\label{fig:4.1_grafica_xvalues}
\end{figure}

In Fig. \ref{fig:4.1_grafica_xvalues} the evolution of concentration with time at different positions ($x=0$, $x=0.1L$, and $x=0.2L$) is represented. It can be observed that the results are very close with maximum differences of $1.39\%$ for the position $x=0.2L$ at time $t=10000s$.

\subsection{Diffusive behaviour}

The second test is designed to evaluate the FFT solution of the diffusion equation in the presence of heterogeneity in the diffusivity of the media. The domain is divided in two equal vertical bands, and the diffusivity in the left part is $D_l=10^{-6}$mm$^2$/s and $D_r=10^{-5}$mm$^2$/s in the right part. No external flux is considered, and the initial condition consists of a concentration of $c_{0,max}=0.98$ in a circle of radius $r=0.3568$mm located in the center of the domain and a concentration of $c_{0,min}=10^{-2}$ out of that circle. The size of the domain is $L_x=1$mm,$L_y=1$mm and is discretized in $N_x=100$, $N_y=100$ voxels. The FE mesh has thr same number of degrees of freedom and contains 20000 elements. \bluecom{Note that due to the discretization in voxels natural in FFT a material point lies in one phase or the other, being the boundary between the two phases stepwise. Contrary, in FE method the boundary can have an arbitrary curvilinear shape using a adaptive meshing. Steps in the FFT boundary are a disadvantage of the method since they can lead to numerical noise. Different solutions have been proposed to limit these problems, as introducing a conformal map to adapt the grid to the microstructure shape \citep{zecevic2021} or using interface voxels with average properties from both phases \citep{KABEL2015168}, used for example to model surface of lattice materials in \cite{LUCARINI2022114223}. However, in this paper, such modifications are not included since the surface behavior is not critical for that study and also because the diffusive nature of the problem naturally aleviate this issue.}

\begin{figure}[H]
	\centering
	\begin{subfigure}{0.3\textwidth}
		\includegraphics[width=\textwidth]{x0.png}
		\caption{$c$ in $x=0.0l$}
		\label{fig:4.2_grafica_x00}
	\end{subfigure}
	\hspace{2mm}
	\begin{subfigure}{0.3\textwidth}
		\includegraphics[width=\textwidth]{x02.png}
		\caption{$c$ in $x=0.2l$}
		\label{fig:4.2_grafica_x02}
	\end{subfigure}	
	\hspace{2mm}
	\begin{subfigure}{0.3\textwidth}
		\includegraphics[width=\textwidth]{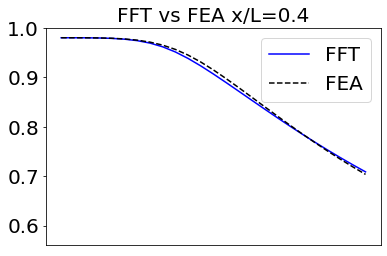}
		\caption{$c$ in $x=0.4l$}
		\label{fig:4.2_grafica_x04}
	\end{subfigure}	
 
	\begin{subfigure}{0.3\textwidth}
		\includegraphics[width=\textwidth]{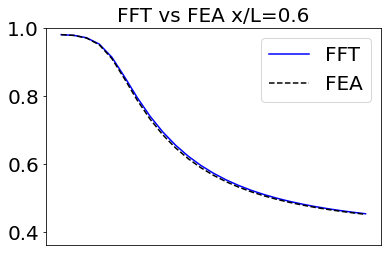}
		\caption{$c$ in $x=0.6l$}
		\label{fig:4.2_grafica_x06}
	\end{subfigure}
	\hspace{2mm}
	\begin{subfigure}{0.3\textwidth}
		\includegraphics[width=\textwidth]{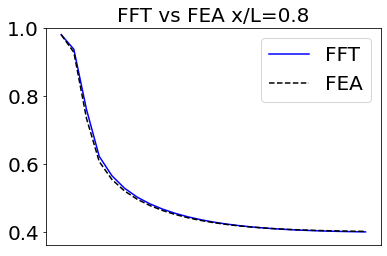}
		\caption{$c$ in $x=0.8l$}
		\label{fig:4.2_grafica_x08}
	\end{subfigure}	
	\hspace{2mm}
	\begin{subfigure}{0.3\textwidth}
		\includegraphics[width=\textwidth]{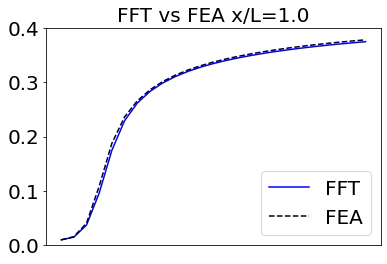}
		\caption{$c$ in $x=1.0l$}
		\label{fig:4.2_grafica_x1}
	\end{subfigure}	
 
	\caption{Concentration ($c$) evolution with time at different points in the plane $y=L/2$ and $x/L=0,0.2,0.4,0.5,0.6,0.8$}
	\label{fig:4.2_grafica_xvalues}
\end{figure}

\begin{figure}[H]
	\centering
	\begin{subfigure}[h]{0.3\linewidth}
        \centering
		\includegraphics[width=\linewidth]{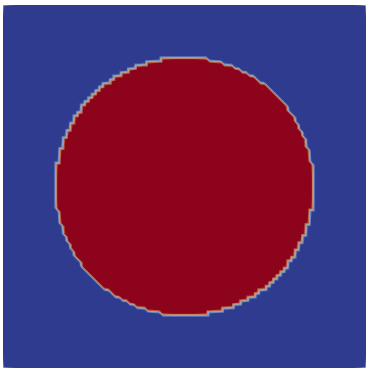}
		\label{fig:4.2_dibujo_t0_fe}
	\end{subfigure} 
    \hspace{2mm} 
	\begin{subfigure}[h]{0.3\linewidth}
        \centering
		\includegraphics[width=\linewidth]{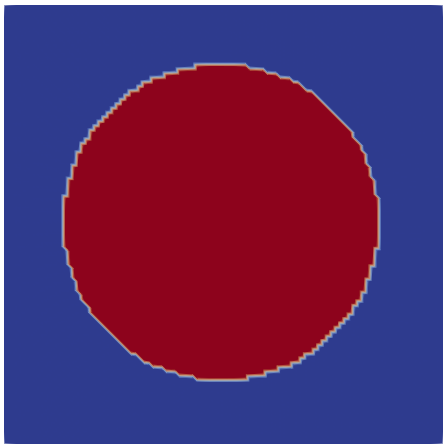}
		\label{fig:4.2_dibujo_t0_fft}
	\end{subfigure} 
    
	\begin{subfigure}[h]{0.3\linewidth}
        \centering
		\includegraphics[width=\linewidth]{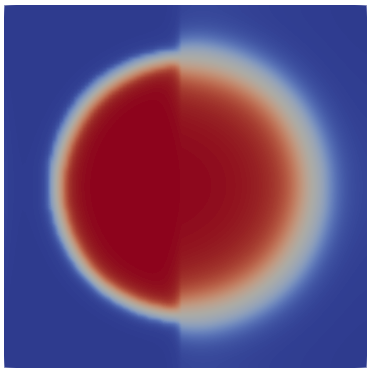}
		\label{fig:4.2_dibujo_t1_fe}
	\end{subfigure} 
    \hspace{2mm} 
	\begin{subfigure}[h]{0.3\linewidth}
        \centering
		\includegraphics[width=\linewidth]{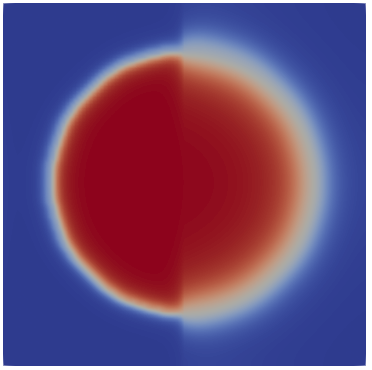}
		\label{fig:4.2_dibujo_t1_fft}
	\end{subfigure} 

	\begin{subfigure}[h]{0.3\linewidth}
        \centering
		\includegraphics[width=\linewidth]{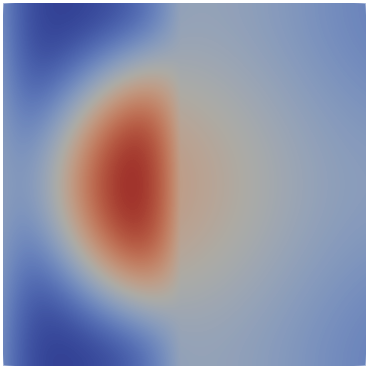}
		\label{fig:4.2_dibujo_t14_fe}
	\end{subfigure} 
    \hspace{2mm} 
	\begin{subfigure}[h]{0.3\linewidth}
        \centering
		\includegraphics[width=\linewidth]{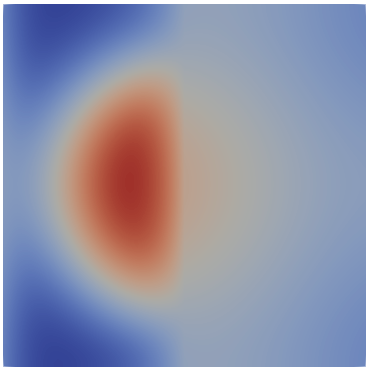}
		\label{fig:4.2_dibujo_t14_fft}
	\end{subfigure} 
    
	\begin{subfigure}[h]{0.3\linewidth}
        \centering
		\includegraphics[width=\linewidth]{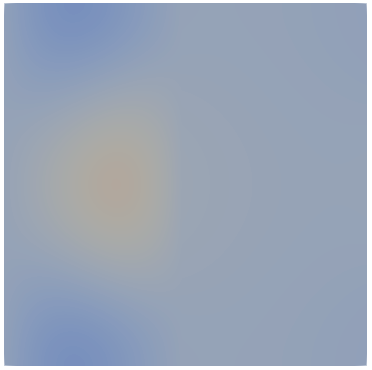}
            \caption{FEA $c$ evolution}
		\label{fig:4.2_dibujo_t49_fe}
	\end{subfigure} 
    \hspace{2mm} 
	\begin{subfigure}[h]{0.3\linewidth}
        \centering
		\includegraphics[width=\linewidth]{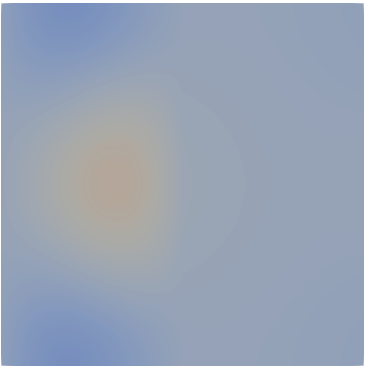}
	\caption{FFT $c$ evolution}	
        \label{fig:4.2_dibujo_t49_fft}
	\end{subfigure} 

        \begin{subfigure}[h]{0.5\linewidth}
        \centering
		\includegraphics[width=\linewidth]{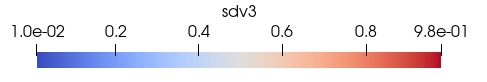}
	\end{subfigure}
 
    \caption{Concentration field ($c$) in a bi-material obtained using FFT and FE solvers. Concentration represented at times $t=0(s)$, $t=1000(s)$, $t=7500(s)$ and $t=25000(s)$.}
    \label{fig:4.2_dibujo_times_fefft} 
\end{figure}

As result of the heterogeneity in the diffusion coefficient, concentration evolves evenly in both halves of the domain, having a faster evolution in the right part because of the higher $D_r$. Fig. \ref{fig:4.2_grafica_xvalues} represents the evolution of the concentration with the time at different positions $x$ in the plane $y=L/2$ for both FE and the FFT algorithm developed. It can be observed that differences are very small, with a maximum difference in time of $c_{dif}=2.94\%$ and average difference during the simulation $\overline{c}_{dif}=0.58\%$ for the position $x=0.2L$ \ref{fig:4.2_grafica_x02} which is the most critical point analyzed. Finally, in Fig. \ref{fig:4.2_dibujo_times_fefft} the concentration map in the domain at different steps is compared, showing again almost identical results for the FFT and FE frameworks.

\subsection{Chemo mechanical induced fracture}\label{example:4.3}
\begin{figure}[H]
	\centering
    \includegraphics[width=0.8\linewidth]{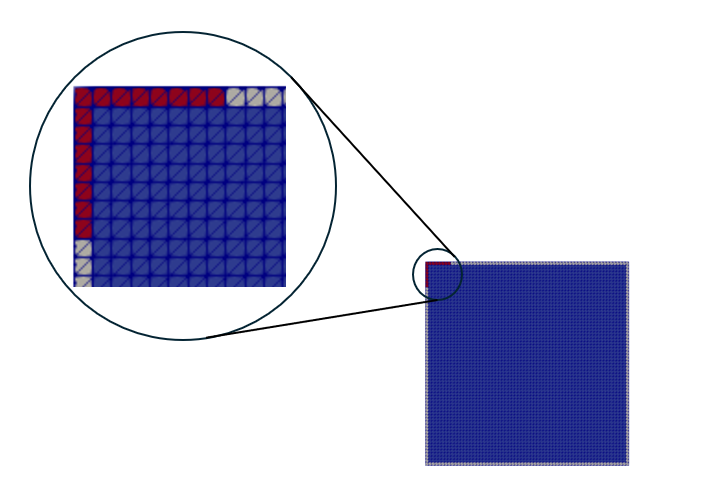}
    \caption{Mesh used in FE formed by triangular elements oriented at 45º}
	\label{fig:mesh_zoom}
\end{figure}

Next benchmark is aimed at evaluating the fully coupled framework in FFT including fracture. The problem consists in a thin plate of a homogeneous material with a concentration flux on its boundaries and free to deform. This concentration flux will generate an heterogeneous concentration distribution in the plate which will induce stresses that will eventually lead to the fracture of the plate.
The plate has dimension $L_x=0.008 \mu$ m, $L_y=0.008 \mu$ m, $L_z=0.000125 \mu$ m discretized in $N_x=64$, $N_y=64$, $N_z=1$ voxels. In FE the mesh has same number of degrees of freedom, resulting in 8192 elements. (Fig. \ref{fig:mesh_zoom}), corresponding to each square voxel of the FFT model two triangles with inclined side at 45$^\circ$.

The material has an elastic modulus of $E=15$Gpa and Poisson's ratio $\nu=0.3$. The fracture toughness is $g_c=2 \ 10^{-3}$Mpa m and the characteristic length $l_c=2.5\ 10^{-4}$ which corresponds to 2 voxels length. The stress threshold is set to $\sigma _{max}=50$Mpa. The flow imposed in the corner of the boundary is $J_{in}=3 \ 10^{-4}$mol/mm$^3$s  while in the rest of the boundaries corresponds to $J_{in}=2\ 10^{-4}$mol/mm$^2$s, for FFT this flux is distributed in an area with $\delta L $ having the value of one voxel (Eq. \ref{eq:delta}). The diffusivity has a value of $D=10^{-9}$mm$^2$/s and the swelling parameter $\Omega =1.3$. 

\begin{figure}[H]
	\centering
    \begin{subfigure}[h]{0.45\linewidth}
		\includegraphics[width=\linewidth]{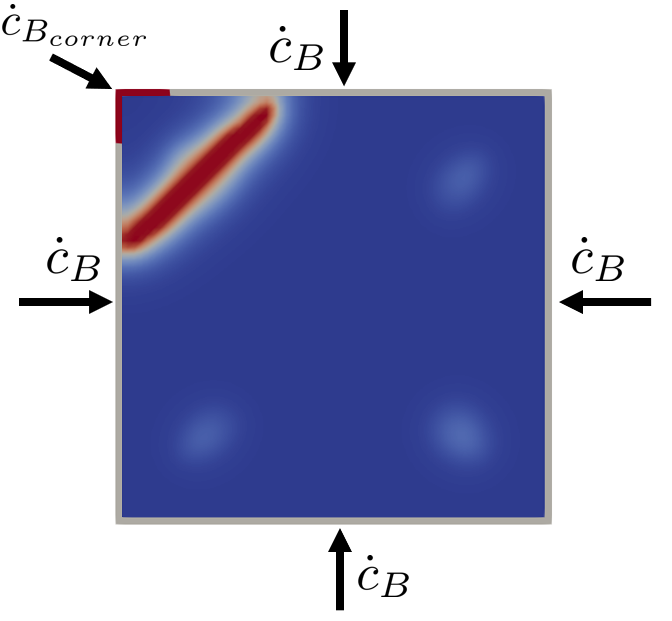}  
    \end{subfigure}
    \begin{subfigure}[h]{0.45\linewidth}
        \includegraphics[width=\linewidth]{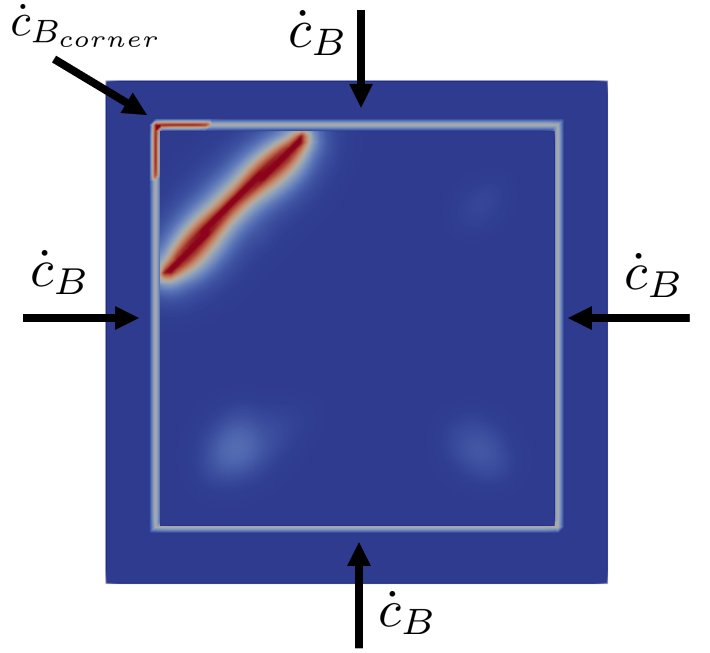}
    \end{subfigure}
    \caption{Flux boundary conditions and resulting crack. Left FE simulation, right FFT results.}
	\label{fig:FE_FFT_corner}
\end{figure}

 To eliminate the symmetry of the problem for an homogeneous material ---which would lead to spurious location of the crack in the multi-sites with same stress concentration--- a higher flux is set in one of the corners, as represented in Fig. \ref{fig:FE_FFT_corner}.

To solve the problem in the FFT framework, the inherent periodic boundary conditions in $c,d$ and $\mathbf{F}$ have to be circumvented. To this aim, a buffer media is introduced around the plate which emulates the empty space, having negligible stiffness ($E_{buffer}=E_{plate}10^{-5}$) to allow the plate to deform freely and very low diffusivity to impede the diffusion out of the physical plate ($D_{buffer}=D_{plate}10^{-5}$). The actual size and shape of the buffer region are represented in Fig. \ref{fig:FE_FFT_corner}. 

\begin{figure}[H]
	\centering

    \begin{subfigure}[h]{0.35\linewidth}
		\includegraphics[width=\linewidth]{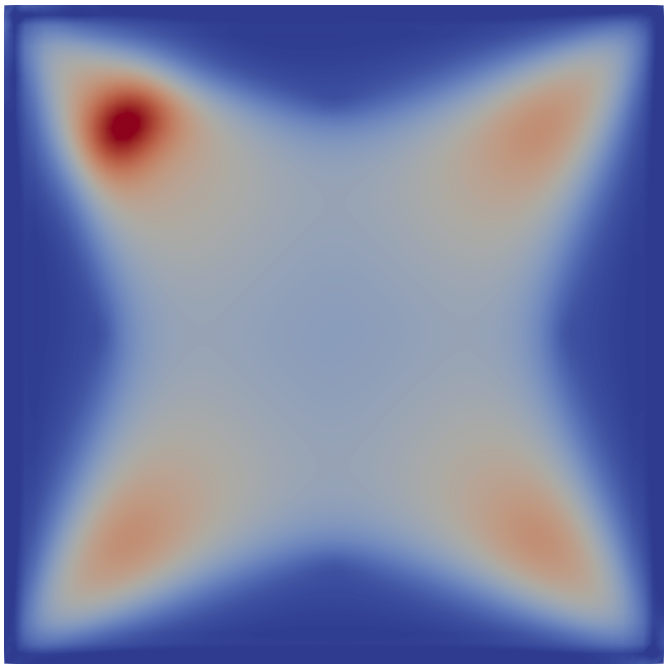}     
    \end{subfigure}
    \hspace{2mm} 
    \begin{subfigure}[h]{0.35\linewidth}
        \includegraphics[width=\linewidth]{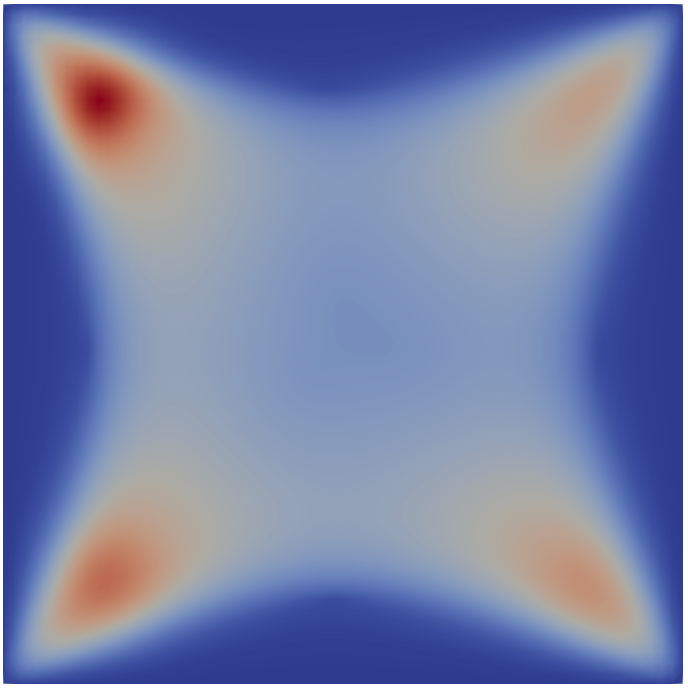}        
    \end{subfigure}

    \begin{subfigure}[h]{0.8\linewidth}
        \includegraphics[width=\linewidth]{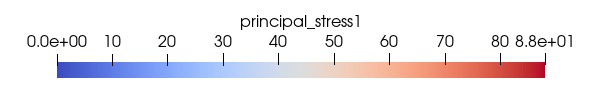}         
    \end{subfigure}
     
    \caption{\bluecom{Principal stress for an upper-left flux before the nucleation of the crack}. Left FE simulation, right FFT results.}
	\label{fig:FE_FFT_stress}
\end{figure}

The stress distribution obtained before the nucleation of a crack is represented in Fig \ref{fig:FE_FFT_stress} for both FE and FFT. It can be observed that stress localizes around the corner where the input flux is maximum at approximately $x=0.8L_x$, $y=0.8L_y$. The result of this stress concentration is the nucleation of a crack. The results of the damage field in both simulations, showing the shape and length of the nucleated crack, are shown in Figs. \ref{fig:FE_FFT_corner} for both the FE and FFT frameworks. It can be observed in Fig. \ref{fig:FE_FFT_corner} that crack position in both solvers is identical, but FFT predicts a thinner crack than FE. The differences between these results are related with the discretization. In FFT, the grid is regular and fields are represented by trigonometrical polynomials. The crack orientation is at -45$^\circ$ and as result, damage field  becomes thicker in this case. 

\vspace{5mm}

\begin{table}[H]
    \begin{center}
       \begin{tabular}{| c | c |} \hline
        Property(units) & Value \\ \hline
        $E_{mat}(Gpa)$ & $15$ \\ \hline
        $E_{void}(Gpa)$ & $1\ 10^{-5} E_{mat}$ \\ \hline
        $nu$ & $0.3$ \\ \hline
        $\sigma_{max}(Mpa)$ & $50$ \\ \hline
        $g_c(Mpa/mm^{1/2})$ & $2\ 10^{-3}$ \\ \hline
        $l_c(mm)$ & $2.5\ 10^{-4}$ \\ \hline
        $\Omega$ & $1.3$ \\ \hline
        $D_{mat}(mm^2/s)$ & $1\ 10^{-9}$ \\ \hline 
        $D_{void}(mm^2/s)$ & $1\ 10^{-5} D_{mat}$ \\ \hline 
        \end{tabular} 
        \caption{Properties of plate for fracture simulation}
        \label{tab:4.3_properties}
    \end{center}
\end{table}

\begin{figure}[H]
	\centering
	\begin{subfigure}[h]{0.3\linewidth}
        \centering
		\includegraphics[width=\linewidth]{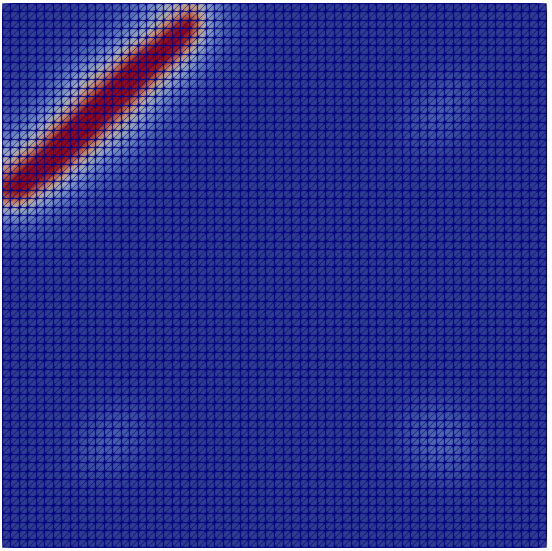}
        \caption{Flux top-left FEA}
		\label{fig:4.3_damage_fe_tl}
	\end{subfigure} 
    \hspace{2mm} 
	\begin{subfigure}[h]{0.3\linewidth}
        \centering
		\includegraphics[width=\linewidth]{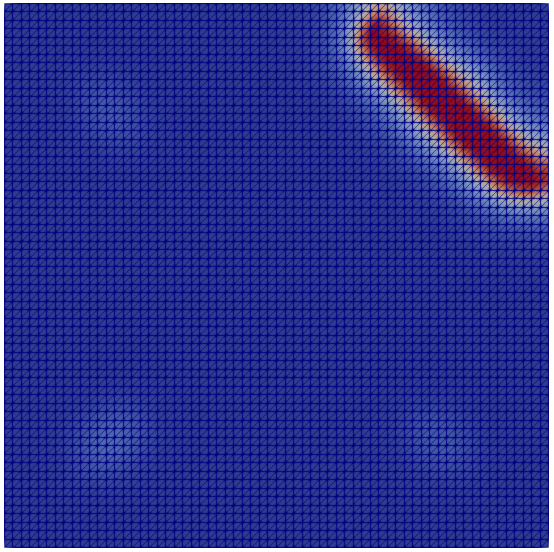}
        \caption{Flux top-right FEA}
		\label{fig:4.3_damage_fe_tr}
	\end{subfigure} 

	\begin{subfigure}[h]{0.3\linewidth}
        \centering
		\includegraphics[width=\linewidth]{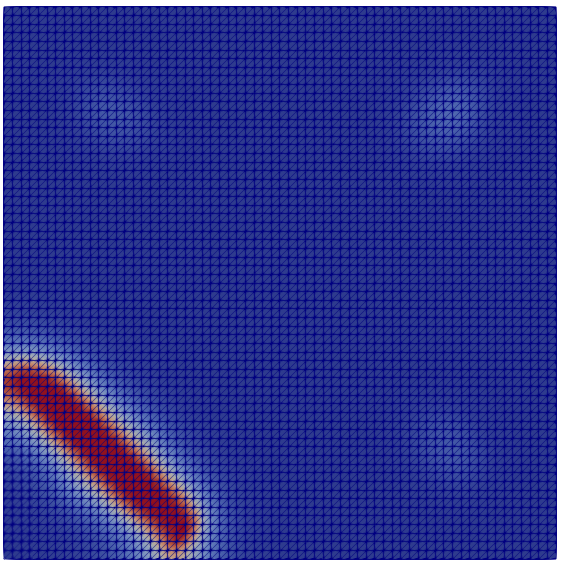}
        \caption{Flux bottom-left FEA}
		\label{fig:4.3_damage_fe_bl}
	\end{subfigure} 
    \hspace{2mm} 
	\begin{subfigure}[h]{0.3\linewidth}
        \centering
		\includegraphics[width=\linewidth]{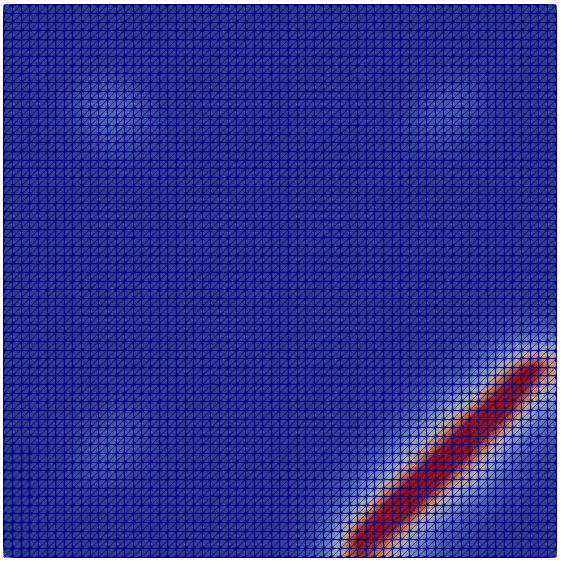}
        \caption{Flux bottom-right FEA}
		\label{fig:4.3_damage_fe_br}
	\end{subfigure} 
 
        \begin{subfigure}[h]{0.5\linewidth}
        \centering
		\includegraphics[width=\linewidth]{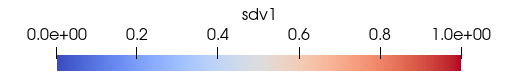}
	\end{subfigure}
 
	\caption{Crack shape result of FE simulations with higher input flux  on the different four corners}
	\label{fig:4.3_damage_fe_all} 
\end{figure}

\begin{figure}[H]
	\centering
	\begin{subfigure}[h]{0.3\linewidth}
        \centering
		\includegraphics[width=\linewidth]{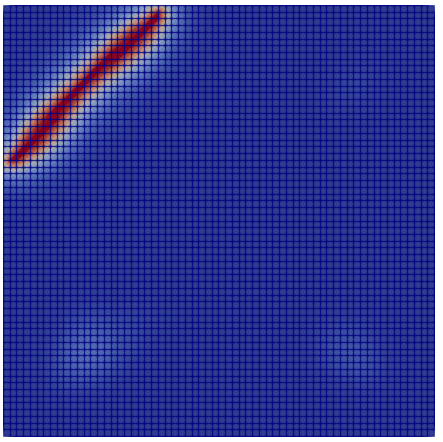}
        \caption{Flux top-left FFT}
		\label{fig:4.3_damage_fft_tl}
	\end{subfigure} 
    \hspace{2mm} 
	\begin{subfigure}[h]{0.3\linewidth}
        \centering
		\includegraphics[width=\linewidth]{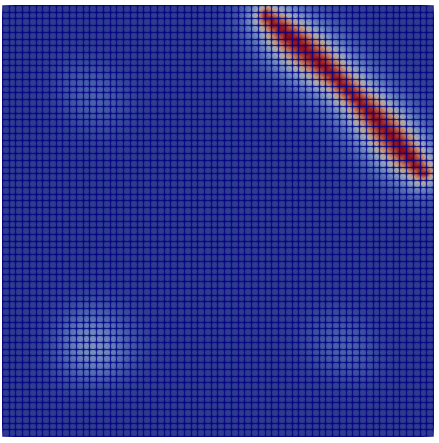}
        \caption{Flux top-right FFT}
		\label{fig:4.3_damage_fft_tr}
	\end{subfigure} 

	\begin{subfigure}[h]{0.3\linewidth}
        \centering
		\includegraphics[width=\linewidth]{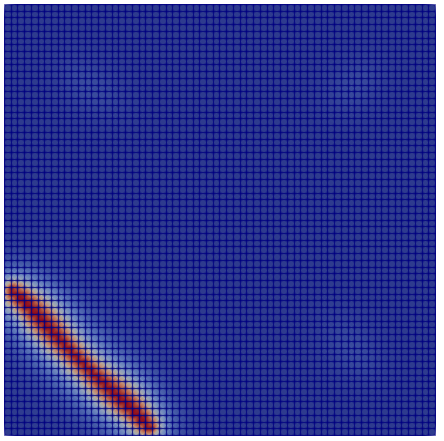}
        \caption{Flux bottom-left FFT}
		\label{fig:4.3_damage_fft_bl}
	\end{subfigure} 
    \hspace{2mm} 
	\begin{subfigure}[h]{0.3\linewidth}
        \centering
		\includegraphics[width=\linewidth]{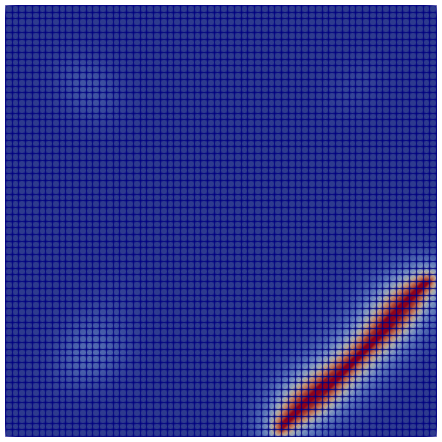}
        \caption{Flux bottom-right FFT}
		\label{fig:4.3_damage_fft_br}
	\end{subfigure} 

        \begin{subfigure}[h]{0.5\linewidth}
        \centering
		\includegraphics[width=\linewidth]{dlegendh.png}
	\end{subfigure} 
 
	\caption{Crack shape result of FFT simulations with higher input flux  on the different four corners}
    \label{fig:4.3_damage_fft_all} 
 
\end{figure}

To further analyze the influence of crack orientation in the shape of the crack, the same problem is repeated in both numerical frameworks changing the corner in which higher flux is imposed. The results in FE and FFT are represented in Figs. \ref{fig:4.3_damage_fe_all} and \ref{fig:4.3_damage_fft_all}. First observation is that cracks appear in the corresponding corner both in FE and FFT. It can be observed FFT results (Fig. \ref{fig:4.3_damage_fft_all}) are almost identical, with rotations of 90, 180 and 270º for the different corner. In all the cases crack shape and thickness are very similar, only small differences can be found in the bottom left square $x=0.2L_x,y=0.2L_y$ caused by the use of a rotated discrete difference scheme (see \cite{willot2015}).

Contrary, FE results are much more affected by crack orientation with respect triangular elements orientation. Simulation with resulting cracks aligned at 45 degrees (Fig. \ref{fig:4.3_damage_fe_all} (b) and (c)) have very similar crack shapes, same that results with cracks aligned at -45º degrees (Fig. \ref{fig:4.3_damage_fe_all} (a) and (d)), but the resulting shape of cracks at 45 and -45 degrees are different. This differences are due to the relative orientation of triangles with respect crack orientation, and are typical effect in FE simulations in similar problems \cite{cervera2022,mandal2019,freddi2022}. The cracks oriented with the mesh have a shape more similar to FFT simulations. 

\section{Simulation of graphite particles during ion intercalation}
In this section, the FFT framework developed and validated against FE in the previous section will be used to predict the fracture of active particles during intercalation. Both 2D and 3D simulations will be presented assuming different  particle shapes and property distributions.

\subsection{Microstructure with random properties distribution}\label{subsec:5.1}
\begin{figure}[H]
    \centering
    \begin{subfigure}[h]{0.5\linewidth}
        \centering
        \includegraphics[width=\textwidth]{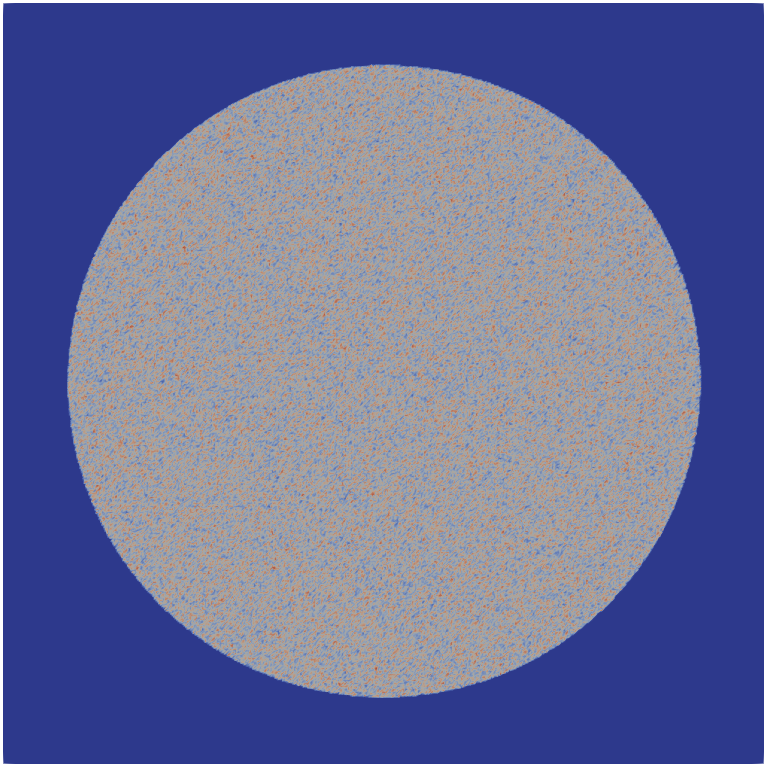}        
    \end{subfigure}
    \begin{subfigure}[h]{0.5\linewidth}
        \centering
        \includegraphics[width=\textwidth]{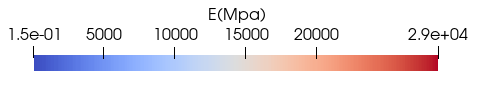}    
    \end{subfigure}

    \caption{Simulation domain for a 2D circular active particle cracking simulation with microscopic property distribution. The external area is occupied by a material with very small stiffness and  diffusivity to emulate empty space. The local value of the Young's modulus of the particle $E$ follows a Gumbell distribution and is represented with a colour scale.}
    \label{fig:5.1_properties}
\end{figure}

The first example consists in a circular particle in which some local properties are set randomly following a Gumbell distribution to consider the microscopic variation of properties in graphite, as proposed in \cite{roque2023}. The full simulation domain is a square of size $L_x=16\mu m$, $L_y=16\mu m$ and the circle is located in the center of the square, having a diameter of $d_p=12.8\mu m$ \ref{fig:5.1_properties}. The domain is discretized in $N_x=512$, $N_y=512$ voxels.

\bluecom{The particle is hyperelastic and follows a Saint-Venant Kirchoff model}, and its average properties are a Young's modulus of $E_{mat}=15$GPa and Poisson's ratio of $\mu=0.3$. The Young's modulus of each element is modified for including heterogeneity through a Gumbell distribution
\begin{equation}\label{eq:Gumbell}
    E_{mat,i}=E_{mat} \log \{1/(1-\omega_i)\}^{(1/m)}
\end{equation}
where $\omega_i$ is a random number in the range $(0,1)$ for each element. The Gumbell distribution exponent is $m=3$. The fracture toughness is $g_c=2\ 10^{-3}$Mpa m and the characteristic length $l_c=1.25\ 10^{-4}$. \bluecom{The average stress threshold defining the particle strength (Eq. \ref{eq:SigDrivingForce}) is set to $\sigma _{max}=100$Mpa, and its local value is set also following a Gumbell distribution (Eq. \ref{eq:Gumbell}) with same parameters.} The swelling parameter is set to $\Omega = 1.3$ and the diffusivity $D=1\ 10^{-9}$mm$^2$/s. All these properties are taken from \cite{roque2023} and their values are discussed in that work. Neumann boundary conditions are considered through the source term $\dot{c}_B$, following Eqs. \eqref{eq:neumann1}-\eqref{eq:neumann2}. In this example, the parameter $\delta L$ defining the area where $\dot{c}_B$ is applied corresponds to 6 voxels, being $\dot{c}_{B}=8 \ 10^{-5}$mol/mm$^3$s$^-1$.

The buffer material in the external area surrounding the particle is modeled with very low stiffness to avoid a constraint in the particle deformation, $E_{void}=E_{mat}\ 10^{-5}$ and with a very small diffusivity $D_{void}=D_{mat} \ 10^{-5}$ to prevent diffusion in that area. All these properties are specified in table \ref{tab:5.1_properties}.

\begin{table}[H]
    \begin{center}
       \begin{tabular}{| c | c |} \hline
        Property(units) & Value \\ \hline
        $E_{mat}$(Gpa) & $15$ \\ \hline
        $m_E$ & $3$ \\ \hline
        $E_{void}$(Gpa) & $1\ 10^{-5} E_{mat}$ \\ \hline
        $nu$ & $0.3$ \\ \hline
        $\sigma_{max}$(Mpa) & $160$ \\ \hline
        $m_{\sigma_{max}}$ & $3$ \\ \hline
        $g_c$(Mpa/mm$^{1/2}$) & $2\ 10^{-3}$ \\ \hline
        $l_c(mm)$ & $2.325\ 10^{-4}$mm \\ \hline
        $\Omega$ & $1.3$ \\ \hline
        $D_{mat}$(mm$^2$/s) & $1\ 10^{-9}$ \\ \hline 
        $D_{void}$(mm$^2$/s) & $1\ 10^{-5} D_{mat}$ \\ \hline 
        \end{tabular} 
        \caption{Properties of a graphite active particle immersed in very compliant and not diffusive material, used in examples \ref{subsec:5.1} and \ref{subsec:5.2}.}
        \label{tab:5.1_properties}
    \end{center}
\end{table}

\begin{figure}[H]
	\centering
	\begin{subfigure}[h]{0.3\linewidth}
		\includegraphics[width=\linewidth]{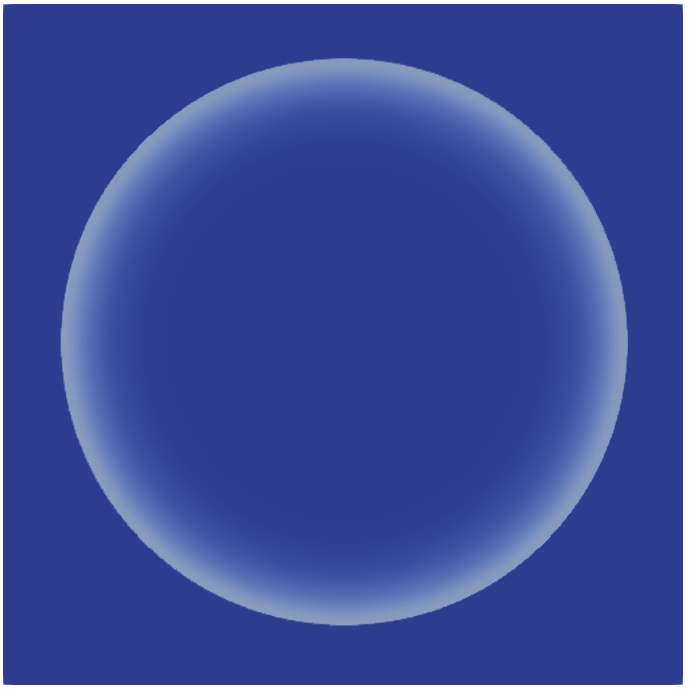}
		\caption{$c$ at time $t =500(s)$}
		\label{fig:5.1_ct500}
	\end{subfigure}
    \hspace{6mm} 
	\begin{subfigure}[h]{0.3\linewidth}
		\includegraphics[width=\linewidth]{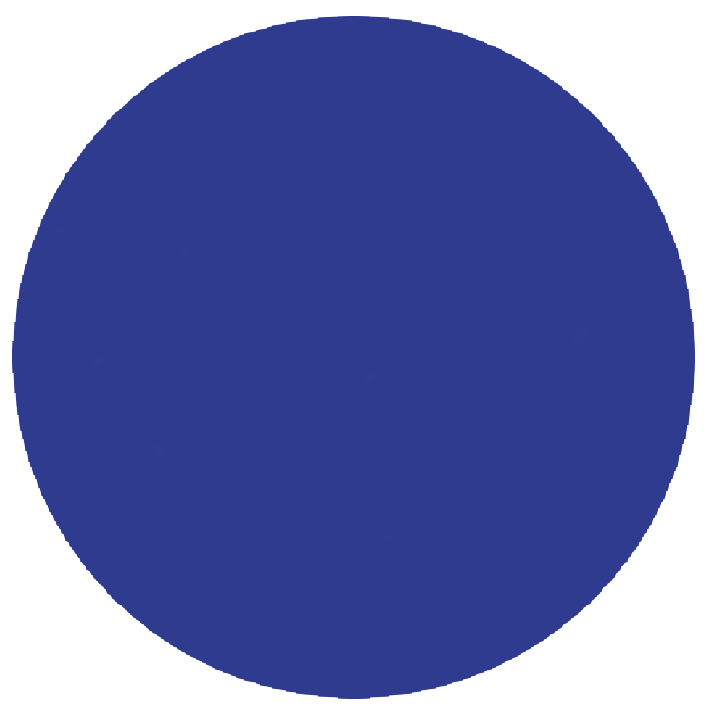}
		\caption{$d$ at time $t =500(s)$}
		\label{fig:5.1_dt500}
	\end{subfigure}		
 
	\begin{subfigure}[h]{0.3\linewidth}
		\includegraphics[width=\linewidth]{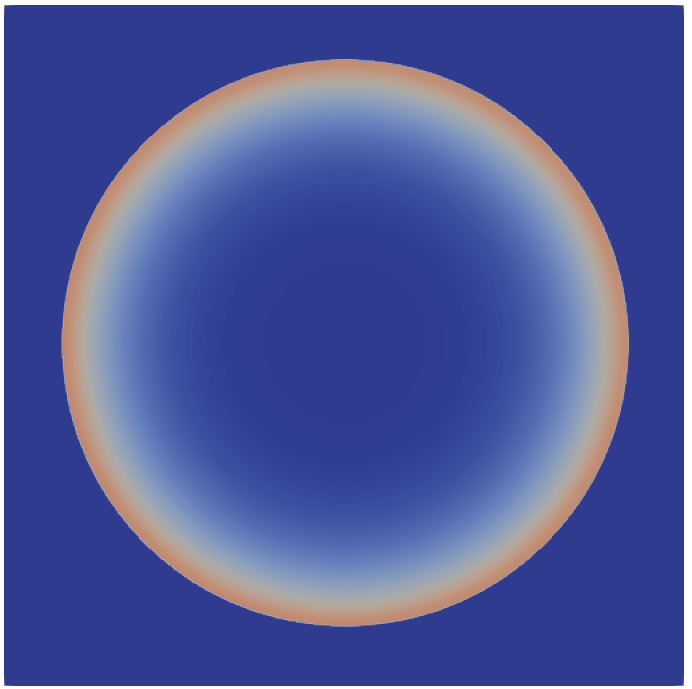}
		\caption{$c$ at time $t =1500(s)$}
		\label{fig:5.1_ct1500}
	\end{subfigure}
    \hspace{6mm}  
	\begin{subfigure}[h]{0.3\linewidth}
		\includegraphics[width=\linewidth]{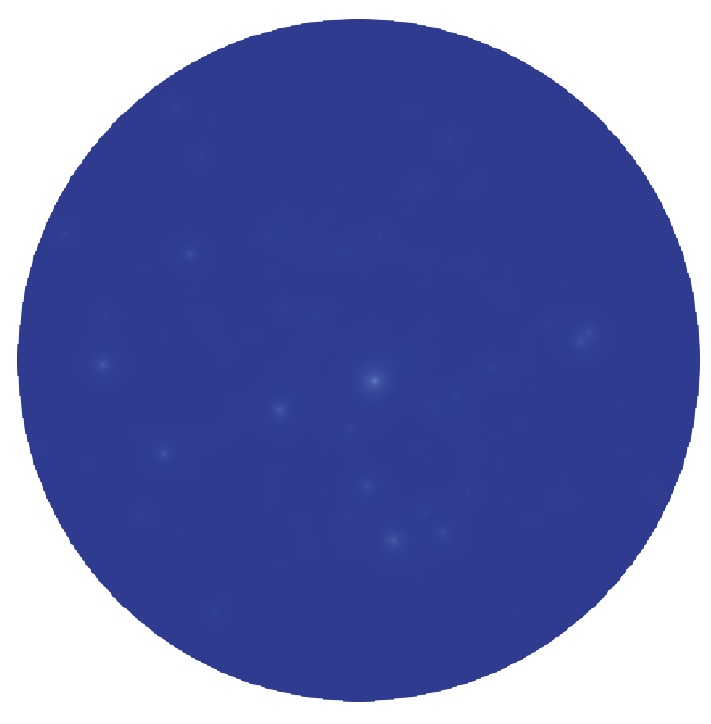}
		\caption{$d$ at time $t =1500(s)$}
		\label{fig:5.1_dt1500}
	\end{subfigure}		
 
 	\begin{subfigure}[h]{0.3\linewidth}
		\includegraphics[width=\linewidth]{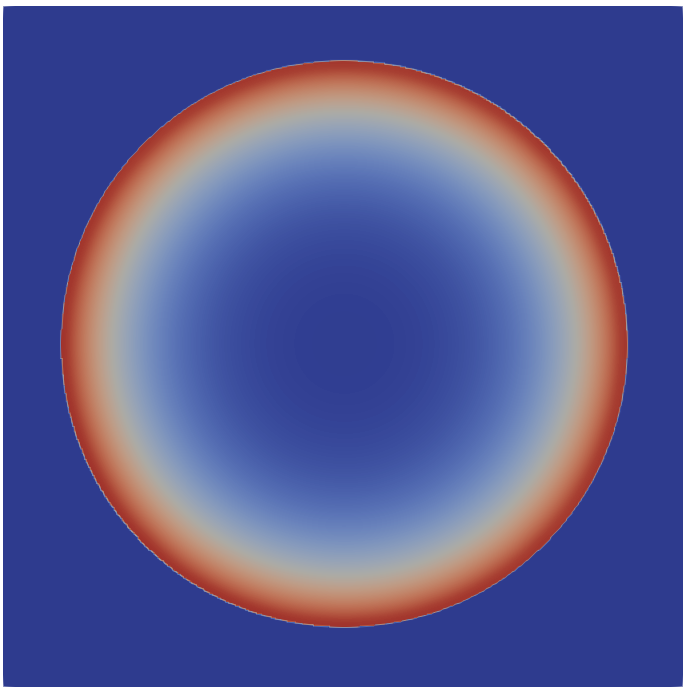}
		\caption{$c$ at time $t =2500(s)$}
		\label{fig:5.1_ct2500}
	\end{subfigure}
    \hspace{6mm} 
	\begin{subfigure}[h]{0.3\linewidth}
		\includegraphics[width=\linewidth]{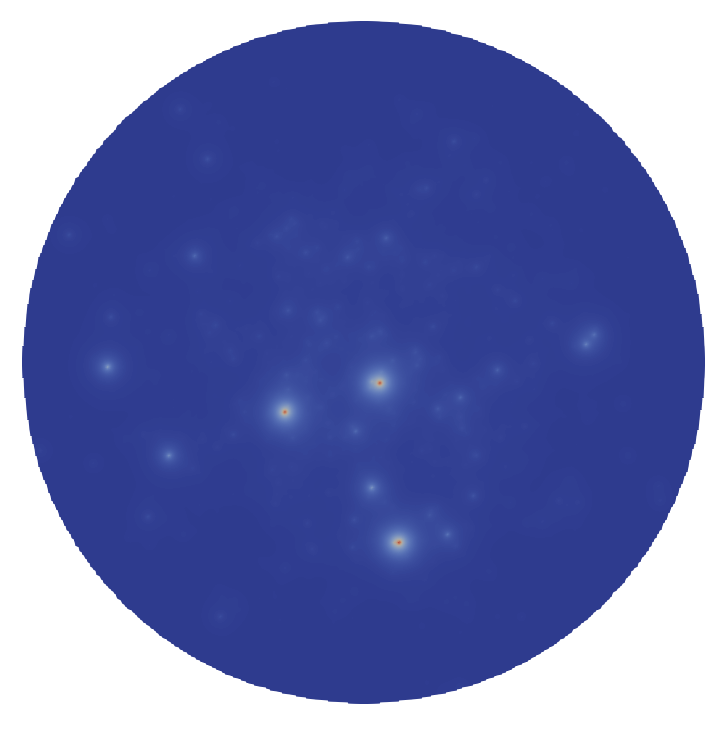}
		\caption{$d$ at time $t =2500(s)$}
		\label{fig:5.1_dt2500}
	\end{subfigure}		
 
	\begin{subfigure}[h]{0.3\linewidth}
		\includegraphics[width=\linewidth]{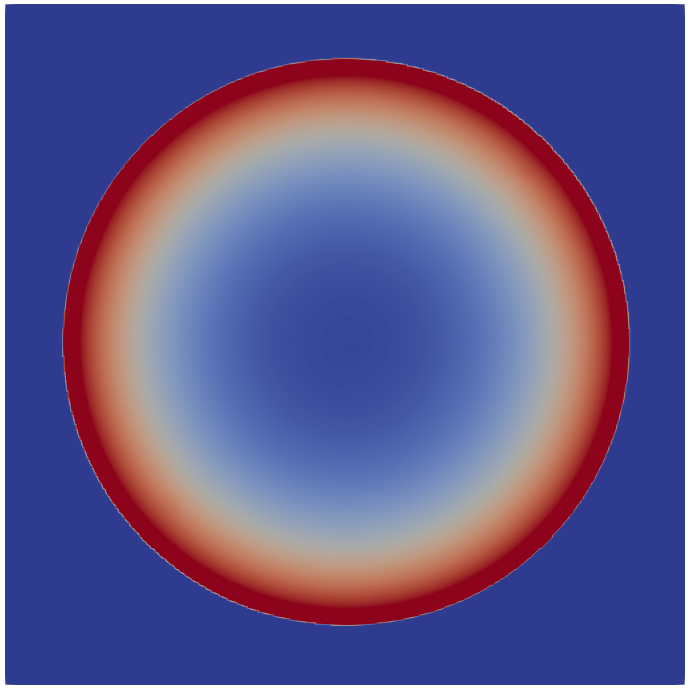}
		\caption{$c$ at time $t =3500(s)$}
		\label{fig:5.1_3500}
	\end{subfigure}
    \hspace{6mm} 
	\begin{subfigure}[h]{0.3\linewidth}
		\includegraphics[width=\linewidth]{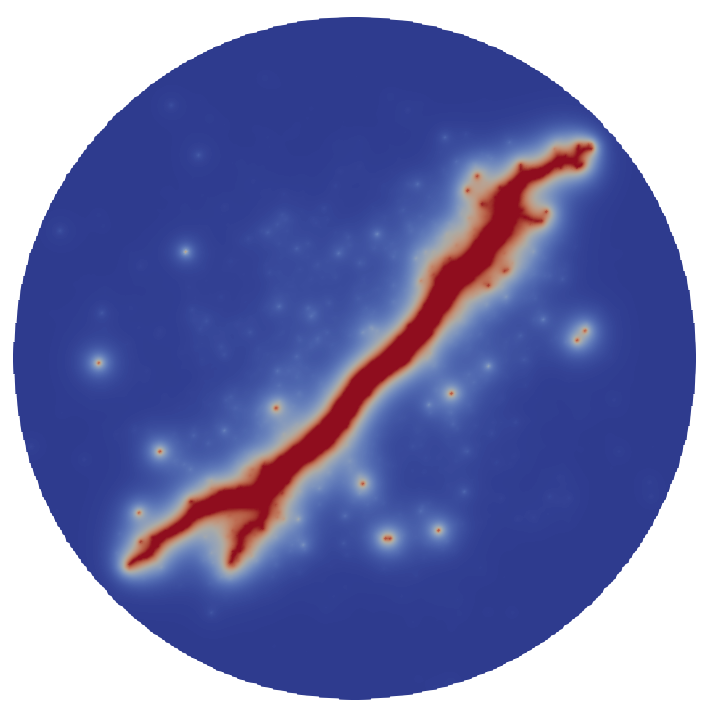}
		\caption{$d$ at time $t =3500(s)$}
		\label{fig:5.1_dt3500}
	\end{subfigure}		

 	\begin{subfigure}[h]{0.4\linewidth}
		\includegraphics[width=\linewidth]{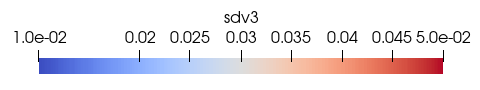}
		\label{fig:5.1_clegend}
	\end{subfigure}
	\begin{subfigure}[h]{0.4\linewidth}
		\includegraphics[width=\linewidth]{dlegendh.png}
		\label{fig:5.1_dlegend}
	\end{subfigure}		
 
	\caption{Fracture process of particle with random distribution of properties. Left column concentration evolution, right column damage evolution. Damage starts to increase in certain critical points, accumulating and leading to the final fracture.}
	\label{fig:5.1_cdtall} 
\end{figure}

The results of this simulation, both concentration and damage fields, are represented in Fig. \ref{fig:5.1_cdtall} for different times. At the initial steps no damage appears due to the threshold $\sigma_{max}$ (fig\ref{fig:5.1_dt500}). The concentration is higher near the particle boundary than in the inner part of the particle, leading to a tensile stress in the central part. The damage starts  in the points in which the tensile strength reaches the local value of $\sigma _{max}$, which happens in random points with higher probability near the particle center. This process continues and several points become damaged fig.\ref{fig:5.1_dt2500}, until the complete fracture of the particle occurs, as shown in Fig. \ref{fig:5.1_dt3500}.

\subsection{3D particle with irregular shape}\label{subsec:5.2}
\begin{figure}[H]
    \centering
    \includegraphics[width=0.5\linewidth]{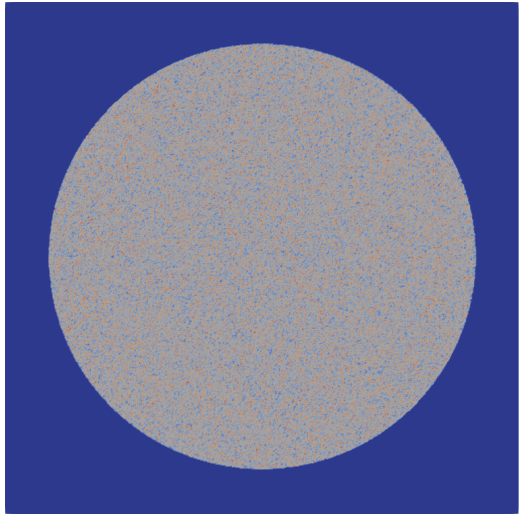}
    \caption{Domains representation of a 3D particle simulation. Active particle immersed in a void material with no diffusivity.}
    \label{fig:5.2_domains}
\end{figure}

The FFT framework developed is now used to simulate the fracture of a 3D particle of graphite with a realistic irregular shape. The RVE is a cube, discretized in $N_{tot}=128^3$ voxels, with a phase surrounding the particle having very small stiffness and conductivity (Fig.\ref{fig:5.2_domains}). An input flux $J_{in}=1.2 \ 10^{-4}$mol/mm$^2$s$^{-1}$ is imposed on the particle boundary being the source term $\dot{c}_{B}=J_{in}/ \delta L$ and $\delta L$ corresponding to one voxel width around the particle. The initial concentration of the particle is $c_0=0.01 $mol/mm$^3$, the diffusivity of the particle is $D= 10^{-9}$mm$^2$/s, the total domain of simulation has a length of $L=15\mu m$ with a particle of irregular shape embedded on its middle part as showed in fig.\ref{fig:5.1_properties}. The material properties are taken for a graphite particle as $E=15$Gpa, $\nu=0.3$ and the fracture toughness is set to $g_c=2\ 10^{-3}$Mpa m, with a characteristic length $l_c=2.325\ 10^{-4}$mm, stress threshold of $\sigma _{max}=160$Mpa  and swelling parameter $\Omega=1.3$. This properties are reviewed in table \ref{tab:5.1_properties}. In this case, the stochastic property distribution in the particle is not considered.

\begin{figure}[H]
	\centering
	\begin{subfigure}[h]{0.35\linewidth}
		\includegraphics[width=\linewidth]{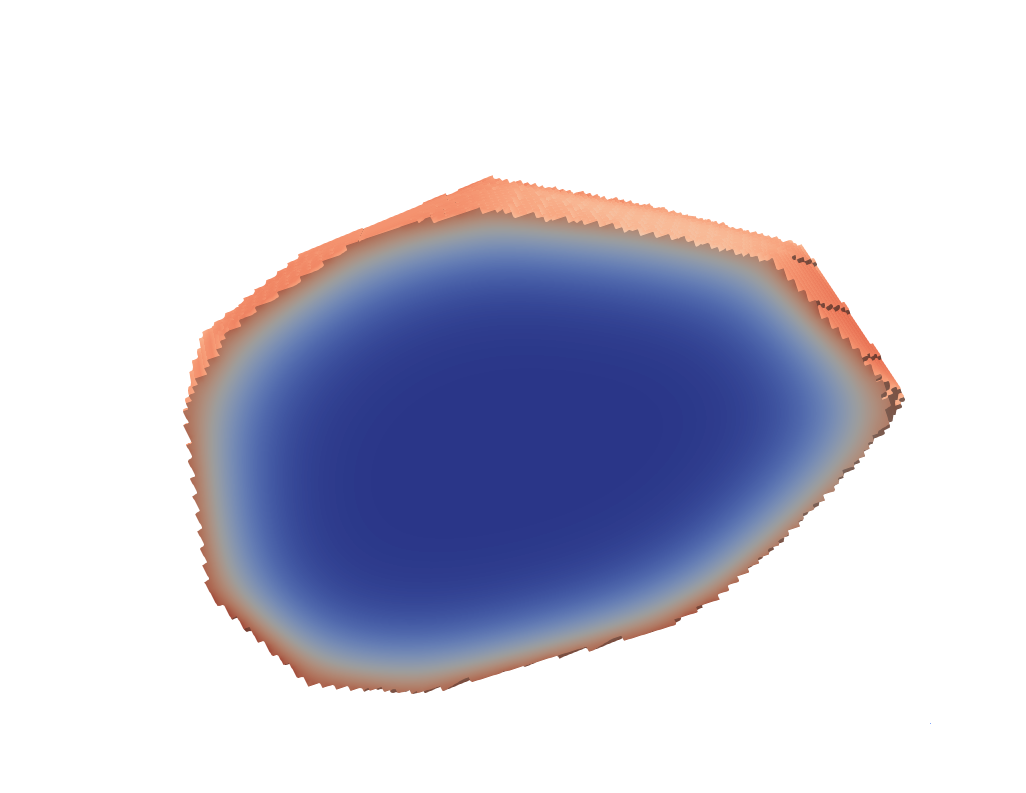}
		\caption{$c$ at $700$s}
		\label{fig:5.2_c_ini}
	\end{subfigure}
	\begin{subfigure}[h]{0.35\linewidth}
		\includegraphics[width=\linewidth]{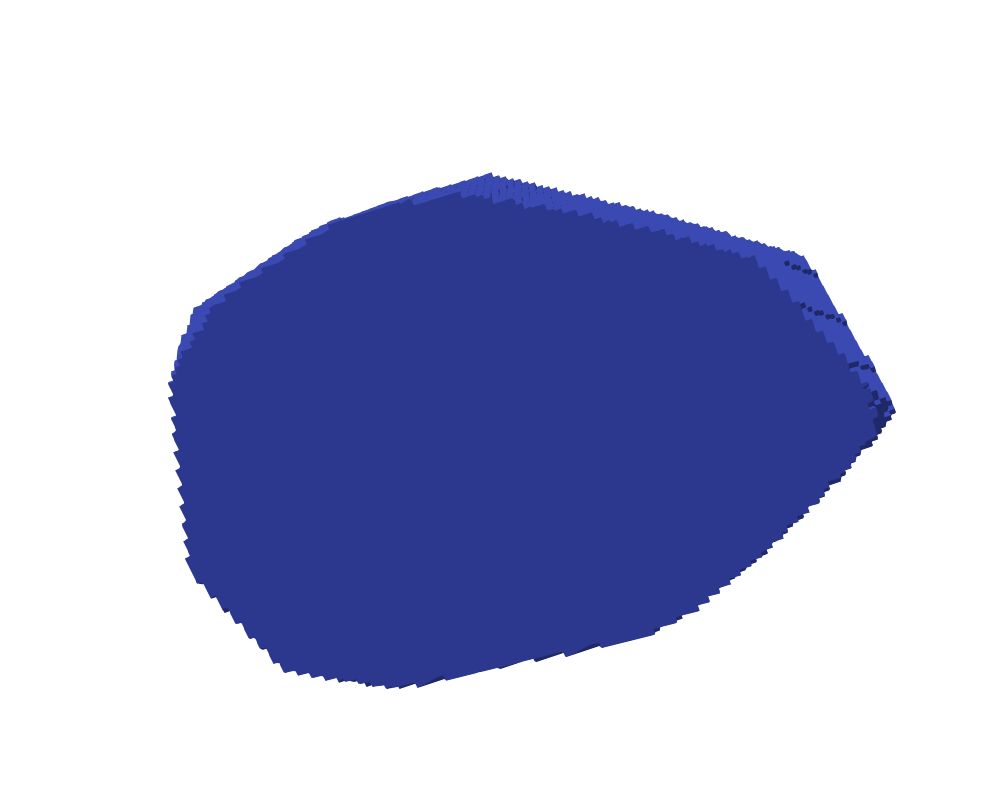}
		\caption{$d$ at $700$s}
		\label{fig:5.2_d_ini}
	\end{subfigure}

	\begin{subfigure}[h]{0.35\linewidth}
		\includegraphics[width=\linewidth]{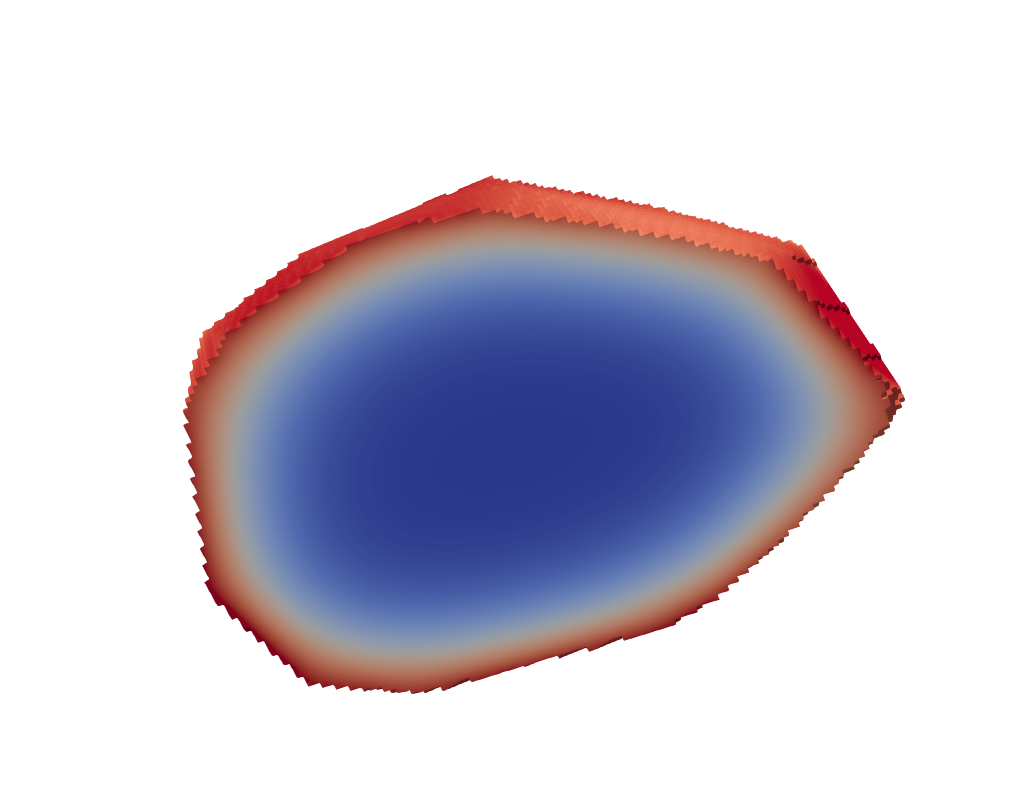}
		\caption{$c$ at $1000$s}
		\label{fig:5.2_c_mid}
	\end{subfigure}
	\begin{subfigure}[h]{0.35\linewidth}
		\includegraphics[width=\linewidth]{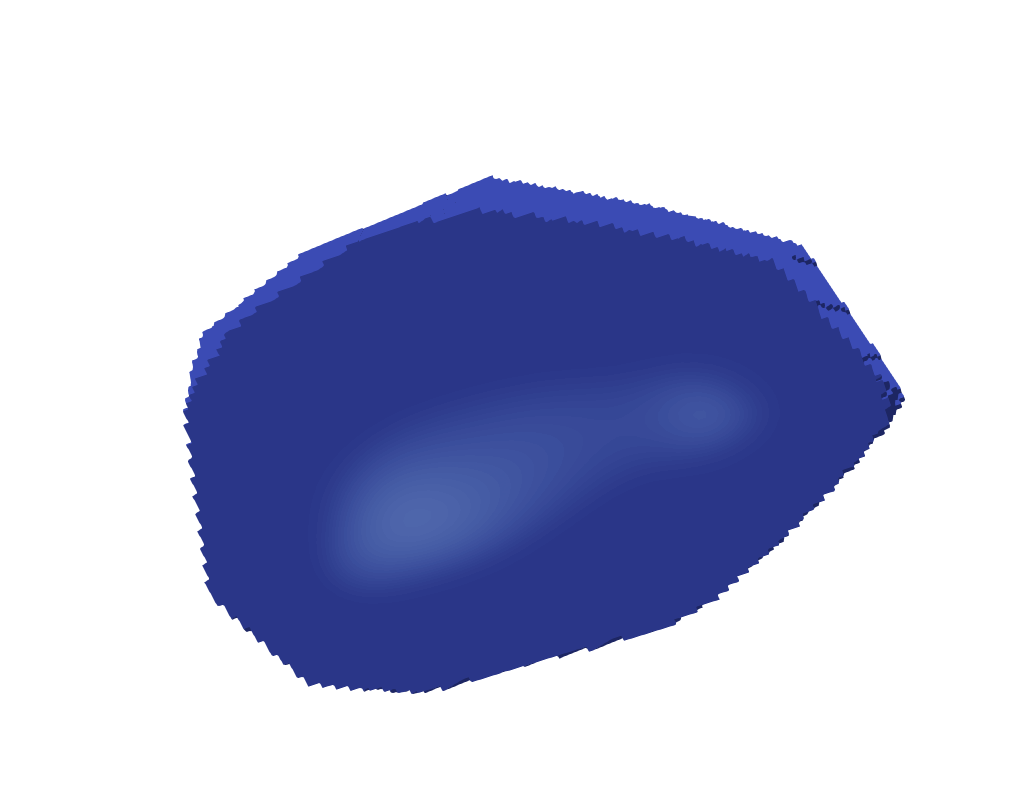}
		\caption{$d$ at $1000$s}
		\label{fig:5.2_d_mid}
	\end{subfigure}		
 
	\begin{subfigure}[h]{0.35\linewidth}
		\includegraphics[width=\linewidth]{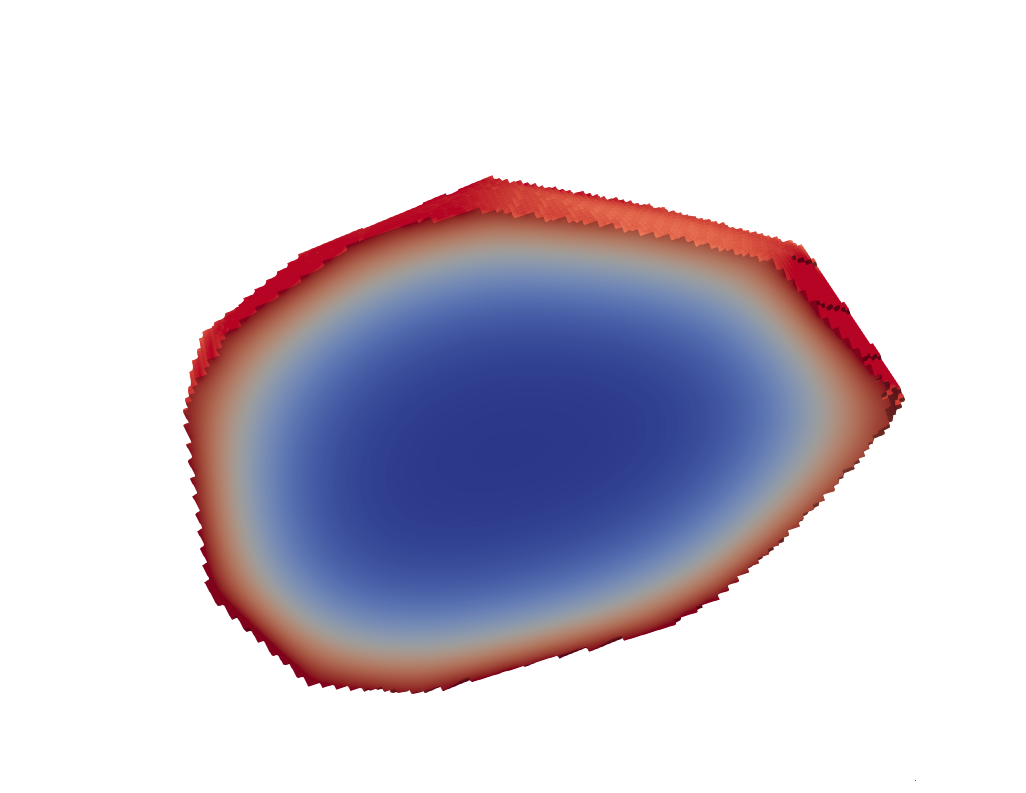}
		\caption{$c$ at $1100$s}
		\label{fig:5.2_c_final}
	\end{subfigure}
	\begin{subfigure}[h]{0.35\linewidth}
		\includegraphics[width=\linewidth]{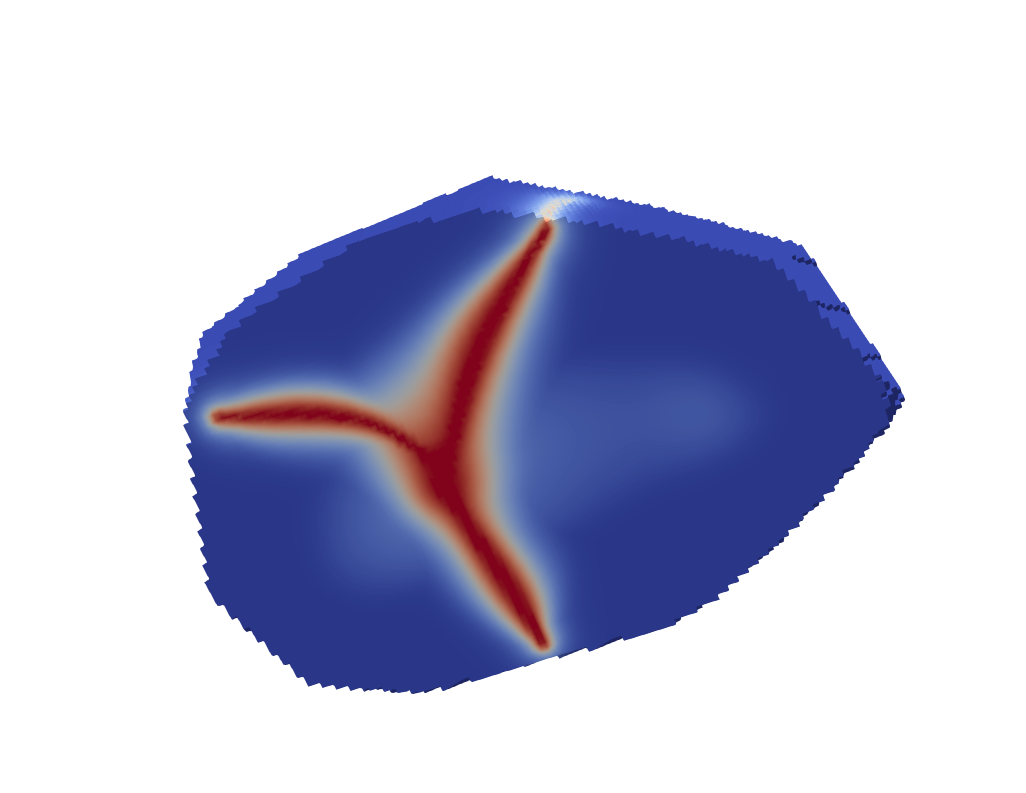}
		\caption{$d$ at $1100$s}
		\label{fig:5.2_d_final}
	\end{subfigure}		
 
	\begin{subfigure}[h]{0.35\linewidth}
		\includegraphics[width=\linewidth]{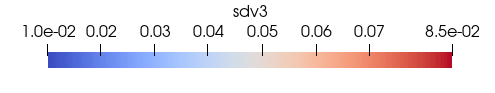}
	\end{subfigure}
	\begin{subfigure}[h]{0.35\linewidth}
		\includegraphics[width=\linewidth]{dlegendh.png}
	\end{subfigure}

	\caption{3D particle fracture process. On the left the concentration evolution due to the input flux increase the concentration in the particle. On the right the damage evolution is represented, at the beginning there is not damage, but when the stress is higher than the stress threshold $\sigma_{max}$ starts to degrade the particle Fig.\ref{fig:5.2_d_mid} which eventually leads the particle cracking Fig.\ref{fig:5.2_d_final}. }
	\label{fig:5.2_fracture_all} 
\end{figure}

The constant flux on the boundary makes the concentration in the external part of the particle to increase, introducing different chemical deformations  inside it. The gradient in the chemical strain generate stresses in the interior of the particle, which increase with the time. During the first steps no damage develops in the particle, Fig.\ref{fig:5.2_d_ini}. When concentration gradients become sufficiently high, the stress reaches the actual strength near the center of the particle  Fig.\ref{fig:5.2_d_mid}, leading to a sudden formation of a crack which cross the particle from surface to surface (Fig.\ref{fig:5.2_d_final}).

Contrary to the case of the circular particle, the complex irregular shape of the 3D particle promotes an heterogeneous stress distribution during the charge. This heterogeneity induces crack formation at a fix position with a well determined shape, preventing the undefined propagation in the case of constant properties and particles with high symmetry. As explained in detail in \cite{roque2023}, the ions diffusion generates high swelling rates in the particle which is well captured by the finite strain formulation. This swelling leads to high tensile stress values in the core of the particle, where the damage starts to increase until the final particle cracking \cite{ryu2018}. The shape of the crack obtained shows an irregular pattern and crosses the particle from side to side, as shown in \cite{harris2010}.
\section{Conclusions}

An FFT based method is developed to simulate chemo-mechanical problems at the microscale including fracture by means of a phase field model. The method involves three fields fully coupled, concentration, deformation gradient and damage and the underlying constitutive model is based on previous works \cite{miehe2015battery} and rooted in a thermodynamically consistent formulation.

In the proposed method the chemical, mechanical and fracture problems are solved in an implicit staggered manner. The mechanical problem is set in a finite strain framework and is solved using Fourier Galerkin method for non-linear problems in finite strains \cite{zeman2017} with mixed control \cite{lucarini20191}. The damage problem is modeled with phase field fracture, using a stress driving force  \cite{miehe2015battery} and solved in Fourier space using a conjugate gradient method with an ad-hoc preconditioner. The chemical problem is modeled with the second Fick's law with physically based chemical potentials. This problem is integrated using backward Euler, and the resulting PDE is solved by Newton-Raphson combined with a conjugate gradient solver for the linearized problems in Fourier space. The three problems are solved under periodic boundary conditions and buffer layers are introduce to break the periodicity and emulate Neumann boundary conditions for incoming mass flux.

This framework is validated against a Finite Element implementation of the same model using the code FEniCS and the results show in general very similar response for both methods. In the case of diffusion, results are almost indistinguishable. It is observed that the emulation of the Neumann boundary conditions in FFT by a source term \bluecom{localized in the surface} provides the same response than actual boundary conditions in FE. The evolution in time of local fields both in a periodic material or an isolated body (considered by a buffer layers in FFT) are almost identical. In the case of a fully coupled problem including mechanical fields and fracture, results of both methods are very similar but small differences are found in the cracks developed. It is observed that in FE crack thickness is influenced by the regular mesh shape, in particular the orientation of triangles with respect the crack. In FFT, the orientation of the crack has less influence, but some asymmetry in the fields is observed due to the use of discrete frequencies \cite{willot2015}.

Finally, the  FFT method is used to solve the fracture of active particles of graphite during ion intercalation, both in 2D and 3D. In this case FE simulations were not performed since the computational cost exceeded the capacity of the computer. The method was able to qualitatively reproduce the shape of the cracks observed in real particles and also provided reasonable values of the flux at which particles break, displaying crack patterns and diffusive behaviour similar to experimental studies \cite{harris2010,roque2023,ryu2018}. The 3D problem on an irregular shaped particle also produce satisfactory results. In this case, the shape and location of the crack is fully determined by its size and shape, contrary to cases with high symmetry and constant properties, where crack can propagate in multiple directions.

As a general conclusion, the method proposed allows to solve coupled problems in complex domains in a robust manner at a relative low computational cost. Even the 3D problem, with more than 2 million of voxels, was solved in a few hours on a single workstation.

\bluecom{In this work, the framework is just applied to the fracture of particles of lithium-ion batteries but could be extended to other applications where a diffusive behavior would be fundamental, like hydrogen embrittlement problems or surface oxidation. The material behavior is given here by finite strain elasticity but could be easily extended to any material law since the Fourier method is material independent and has indeed been applied to many different constitutive equations. As examples, an elastoplastic material law can be used to account for plastification of particles before damage. Moreover, the method would potentially be used to model the lithiation in a full electrode simulation taking advantage of the low computational cost of the proposed FFT framework.}


\end{document}